\documentclass{myarticle}

\usepackage{float}
\usepackage{amssymb}

\usepackage{amsmath} 
\usepackage{amssymb} 
\usepackage{undertilde} 

\newcommand{\norm}[1]{{\| #1 \|}}

\newcommand{\ua}{\underline{a}}
\newcommand{\ub}{\underline{b}}

\newcommand{\uf}{\underline{f}}
\newcommand{\ug}{\underline{g}}

\newcommand{\up}{\underline{p}}
\newcommand{\uq}{\underline{q}}

\newcommand{\ux}{\underline{x}}

\newcommand{\uz}{\underline{z}}

\newcommand{\uC}{\underline{C}}

\newcommand{\uI}{\underline{I}}

\newcommand{\uP}{\underline{P}}

\newcommand{\uS}{\underline{S}}

\newcommand{\uuA}{\underline{\underline{A}}}
\newcommand{\uuB}{\underline{\underline{B}}}
\newcommand{\uuC}{\underline{\underline{C}}}

\newcommand{\uuI}{\underline{\underline{I}}}

\newcommand{\uuK}{\underline{\underline{K}}}

\newcommand{\uuS}{\underline{\underline{S}}}

\newcommand{\utA}{\utilde{A}}
\newcommand{\utB}{\utilde{B}}
\newcommand{\utC}{\utilde{C}}

\newcommand{\utI}{\utilde{I}}

\newcommand{\utK}{\utilde{K}}

\newcommand{\utS}{\utilde{S}}

\newcommand{\ualpha}{\underline{\alpha}}

\newcommand{\uuDelta}{\underline{\underline{\Delta}}}

\newcommand{\cC}{\mathcal{C}}

\newcommand{\cF}{\mathcal{F}}

\newcommand{\cS}{\mathcal{S}}

\newcommand{\bbC}{\mathbb{C}}

\newcommand{\bbI}{\mathbb{I}}

\newcommand{\bbN}{\mathbb{N}}

\newcommand{\bbR}{\mathbb{R}}

\newcommand{\bbZ}{\mathbb{Z}}

\newcommand{\rmd}{\mathrm{d}}

\newcommand{\rmK}{\mathrm{K}}

\newcommand{\rmS}{\mathrm{S}}

\begin{document}

\title{On the generation of periodic discrete structures with identical two-point correlation}
\author[a,b]{Mauricio Fernández\footnote{M. Fernández - \email{fernandez@cps.tu-darmstadt.de}}}
\author[a]{Felix Fritzen\footnote{F. Fritzen - \email{fritzen@mechbau.uni-stuttgart.de}}}
\affil[a]{Data Analytics in Engineering, Stuttgart Center for Simulation Science, Institute of Applied Mechanics (CE), University of Stuttgart, Pfaffenwaldring 7, 70569 Stuttgart, Germany}
\affil[b]{Cyber-Physical Simulation Group, Department of Mechanical Engineering, Technische Universität Darmstadt, Dolivostraße 15, 64293 Darmstadt, Germany}
\maketitle

\begin{abstract}
Strategies for the generation of periodic discrete structures with identical two-point correlation are developed. Starting from a pair of root structures, which are not related by translation, phase inversion or axis reflections, child structures of arbitrary resolution (i.e., pixel or voxel numbers) and number of phases (i.e., material phases/species) can be generated by means of trivial embedding based phase extension, application of kernels and/or phase coalescence, such that the generated structures inherit the two-point-correlation equivalence. Proofs of the inheritance property are provided by means of the Discrete Fourier Transform theory. A Python 3 implementation of the results is offered by the authors through the Github repository \url{https://github.com/DataAnalyticsEngineering/EQ2PC} in order to make the provided results reproducible and useful for all interested readers. Examples for the generation of structures are demonstrated, together with applications in the homogenization theory of periodic media. 

\end{abstract}

\tableofcontents


\section{Introduction}

In several fields of science, discrete structures in one, two, three and higher dimension are essential for the investigation of physical phenomena. For example, two-dimensional images are digitized and represented through matrices, which in this work are considered as two-dimensional structures. Each discrete color in the image can, e.g., be considered as a phase, such that a structure can contain a low or high number of phases. Of course, images ($n$-phase two-dimensional structures) have a wide range of applications. Images are, e.g., used to characterize microstructures in materials science with respect to their evolution or physical properties, see, e.g., \cite{Ditta2020} or \cite{Lissner2019}, and to train modern machine learning approaches in face-recognition, autonomous driving and many other fields, see, e.g., \cite{Iranmanesh2020} or \cite{Sarcinelli2019}. Via computed tomography voxelized three-dimensional information of bodies can be gathered in order to represent three-dimensional structures. The can improve diagnostics in medicine, see, e.g., \cite{Hsieh2009} or \cite{Mahesh2011}, and they can enable in-situ examinations of the evolution of mechanical structures, see, e.g., \cite{Patterson2015} or \cite{Loffl2019}. Space-time data can, in principle, be stored in four-dimensional arrays. 

In terms of data analysis of $n$-phase multi-dimensional structures, the most obvious and simple statistical descriptor is the volume fraction of each phase, which is referred to as the one-point-correlation. For instance, consider a purely black-and-white two-dimensional image of size $4 \times 3$ pixels, where 5 pixels are black. Then the volume fraction of the phase "black" is 5/12 and that of "white" is 7/12. Depending on the quantity of interest, higher-order correlations of the phases may not only offer better insight into the structure, but the can also yield more accurate predictions of the quantity of interest. 

In materials science, the effective behavior of complex heterogeneous materials is dominated by the local material behavior and its structural arrangement at the microscopic level, see, e.g., \cite{Torquato2002}. Therefore, models approximating the effective or homogenized material law and statistical bounds are usually built upon the one-point-correlation of a microscopic representative volume element, at the very least, see, e.g., \cite{Voigt1910}, \cite{Reuss1929}. Here, the representative volume element is assumed as periodic. More accurate models and bounds take into consideration the two-point correlation (2PC) and even multiple-point correlations of the periodic microstructure, see, e.g., \cite{Willis1977}, \cite{Willis1981} or \cite{Drugan2016a}. But a central question arises:  What if two or more structures possess the exact same 2PC? This question has not only problematic implications for the deterministic prediction of models using the 2PC, see, e.g., \cite{Fast2011}, but also, e.g., for structure reconstruction algorithms based on the 2PC, see, e.g., \cite{Yeong1998} and \cite{Fullwood2008b}. 

For non-periodic structures of arbitrary dimensions, \cite{Chubb2000} have shown that the dipole-histogram, which corresponds to the 2PC, is a unique representation of a discrete structure. This implies that if two structures differ, then so do their dipole histograms, such that no group of structures with identical dipole histrograms can exist. For periodic structures the situation is different. In \cite{Jiao2010a} and \cite{Jiao2010b} it has been proven that structures with identical 2PC exist and examples are provided. This implies that the claim of \cite{Fullwood2008b}, that the 2PC uniquely determine a periodic structure, is wrong, as discussed in \cite{Jiao2010b}. A numerical approach for the approximation of the number of structures associated with a given 2PC is investigated in \cite{Gommes2012a} and \cite{Gommes2012b}.

\begin{figure}[H]
\centering
\includegraphics[width=0.9\textwidth]{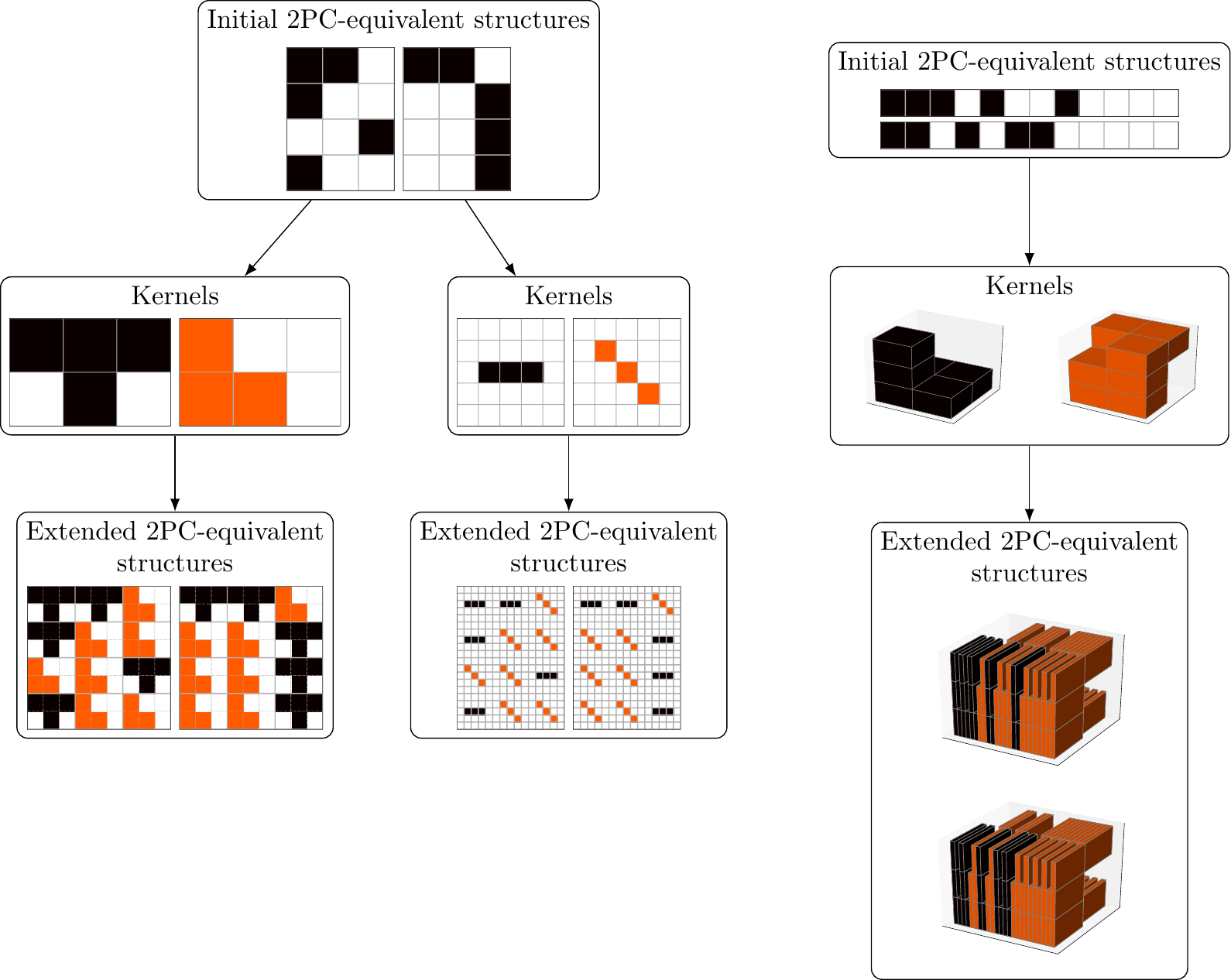}
\caption{Graphical summary of the generation of 2PC-equivalent structures based on a phase extension and kernel application on given root 2PC-equivalent structures: (left) two-dimensional examples; (right) one- to three-dimensional example.}
\label{fig_intro}
\end{figure}

The present work aims, compared to \cite{Jiao2010a} and \cite{Jiao2010b}, at a more constructive generation of periodic discrete structures with identical 2PC starting from given root structures with identical 2PC. The present work relies of the comparison of all $(n-1)n/2$ independent 2PCs and discusses some issues related to the work of \cite{Niezgoda2008}, where it is claimed that only $(n-1)$ 2PCs should suffice. Based on the results of the present work, the statement of \cite{Niezgoda2008} does not hold in the general case, to the very best of our knowledge. This finding can be relevant for many works relying on \cite{Niezgoda2008}, see, e.g., \cite{Panchal2013a}, \cite{Miyazawa2019} or \cite{Ashton2020}. In the present work, three main operations are investigated: phase extension, kernel-based extension and phase coalescence. It is proven that these three basic operations conserve the 2PC-equivalence of the given structures, such that an infinite number of 2PC-equivalent structures can be generated starting from given 2PC-equivalent root structures, as graphically summarized in \autoref{fig_intro}.

Hereby, based on appropriate kernels, structures can be purposely designed for specific applications, as e.g., fiber-, particle-reinforced and polycrystalline materials with arbitrary number of phases, see \autoref{fig_intro} (left). Further, the present work shows that, additionally, low-dimensional 2PC-equivalent structures can immediately be used in order to generate 2PC-equivalent structures of arbitrary dimensions and shapes, see \autoref{fig_intro} (right). The basic operations can be combined and repeated as needed in order to generate 2PC-structures of arbitrary number of phases, dimensions and shapes. In order to make the results of the present work as transparent and useful as possible, a Python 3 implementation is offered by the authors through the Github repository \cite{Fernandez2020eq2pc_github}
\begin{center}
	\url{https://github.com/DataAnalyticsEngineering/EQ2PC}
\end{center}
Further, the present work explores the influence of 2PC-equivalent structures on the effective conductivity behavior of two-dimensional periodic materials. The effective conductivity of the structures is computed based on the FANS approach of \cite{Leuschner2018a}. The present work shows that 2PC-equivalent structures exist, which show a deviation of 16\% in the effective conductivity matrix. This implies that for specific structures homogenization schemes relying solely on the 2PC may need further development, i.e., enrichment, in order to preserve prescribed accuracy. The provided root structures and derived ones may be used as benchmark problems for new homogenization schemes in the future. 

The manuscript is organized as follows: In \autoref{sec_2pc}, $n$-phase discrete periodic structures a defined through multi-dimensional arrays, the two-point correlation and the notion of 2PC-equivalence are specified, and the basic operations preserving 2PC-equivalence are proven based on DFT theory. Examples for the generation of 2PC-equivalent structures and the investigation of the deviation in the effective conductivity are demonstrated in \autoref{sec_examples_applications}. The manuscript ends with conclusions in \autoref{sec_conclusions}. The appendices \autoref{app_proofs}, \autoref{app_niezgoda_counterexample} and \autoref{app_software} offer auxiliary proofs, a discussion with counterexamples on the work of \cite{Niezgoda2008} and details on the provided Python 3 implementation of the results of the present work, respectively. 

\paragraph{Notation.} The sets of natural, integer, real and complex numbers are denoted as $\bbN$, $\bbZ$, $\bbR$ and $\bbC$, respectively. The set of natural numbers including the 0 is addressed as $\bbN_0$. The symbol $i$ is reserved for the complex unit, i.e., $i^2 = -1$. Vectors $\ua \in \bbC^P$ of dimension $P$ are addressed through their explicit comma-separated components as $\ua = (a_0,a_1,\dots,a_{P-1})$. Matrices are denoted by double-underlined characters, e.g., $\uuA$. We address the array dimensionality $D \geq 2$ of a number set through a vector of corresponding length in the superscript, e.g., for $D=3$, $\uP = (2,3,4) \in \bbN^3$, $\bbC^{\uP} = \bbC^{2 \times 3 \times 4}$ holds. Arrays of dimensionality $D \geq 3$ are addressed as $\utA$. Element-wise multiplication is denoted by the Hadamard product $\ua \odot \ub$, $\uuA \odot \uuB$ and $\utA \odot \utB$. Complex conjugation is denoted as $\overline{\utA}$, at what this is carried out element-wise in the array. We define the set $\bbI = \{0,1\}$ and its corresponding extensions $\bbI^P$, $\bbI^{(P_1,P_2)}$ and $\bbI^{\uP}$ with $\uP \in \bbN^D$ for vectors, matrices and $D$-dimensional arrays having components being 0 or 1. Indicator arrays are simply addressed as indicators $\utI_\alpha \in \bbI^{\uP}$. Hereby, the symbol $\alpha$ in the subscript is reserved for phase indices in an $n$-phase structure, i.e., $\alpha \in \{1,\dots,n\}$. The symbols $p$ and $q$ are reserved for 0-based components / position indices, i.e., $I_{\alpha,\up}$ denotes the component at position $\up = (p_1,\dots,p_D) \in \bbN_0^D$ of the indicator $\utI_{\alpha} \in \bbI^{\uP}$ of phase $\alpha$ with $\uP \in \bbN^D$, $p_d \in \{0,\dots,P_d-1\}$ and $d \in \{1,\dots,D\}$. 
\section{Periodic structures and 2-point correlation}
\label{sec_2pc}

\subsection{Preliminaries}

\begin{figure}[H]
\centering
\includegraphics[scale=1]{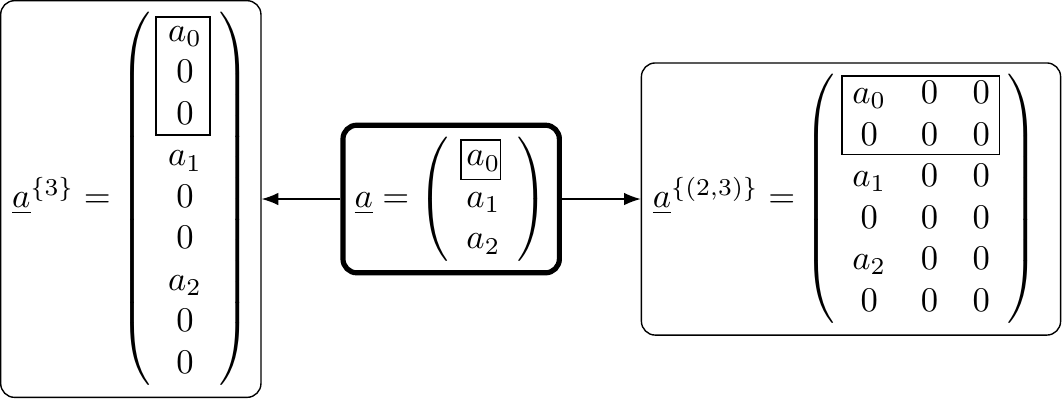}
\caption{Example for the trivial embedding for a vector $\ua \in \bbC^3$ (center): $\ua^{\{3\}} \in \bbC^{9}$ (left); $\ua^{\{(2,3)\}} \in \bbC^{(6,3)}$.}
\label{fig_emb_1d}
\end{figure}

\begin{figure}[H]
\centering
\includegraphics[scale=1]{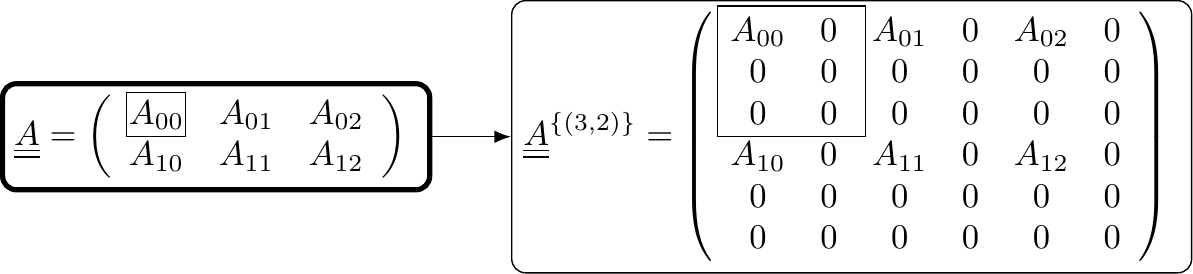}
\caption{Example for the trivial embedding for a matrix $\uuA \in \bbC^{(2,3)}$ (left): $\uuA^{\{(3,2)\}} \in \bbC^{(6,6)}$ (right).}
\label{fig_emb_2d}
\end{figure}

Consider a $D$-dimensional array $\utA \in \bbC^{\uP}$, $\uP = (P_1,\dots,P_D) \in \bbN^D$ with components $A_{\up} = A_{p_1 \dots p_D}$ with position indices $p_r \in \{0,\dots,P_r-1\}$ for $r \in \{1,\dots,D\}$. For an array $\utA \in \bbC^{\uP}$ we define the trivial embedding $\utA^{\{\uz\}}$ with $\uz \in \bbN^N$ through its components as
\begin{eqnarray}
	N \leq D : 
		&&
		\utA^{\{\uz\}} \in \bbC^{\uP'}
		\ , \quad
		\uP' = (z_1P_1,\dots,z_NP_N,P_{N+1},\dots,P_D)
		\nonumber
		\\
		&&
		A^{\{\uz\}}_{p_1 \dots p_N p_{N+1} \dots p_D}
		=
		\begin{cases}
		A_{q_1 \dots q_N p_{N+1} \dots p_D} & p_r = z_r q_r, r \in \{1,\dots,N\} \\
		0 & \text{else}
		\end{cases}
		\label{eq_zp_def1}
		\\
	N > D : 
		&&
		\utA^{\{\uz\}} \in \bbC^{\uP'}
		\ , \quad
		\uP' = (z_1P_1,\dots,z_DP_D,z_{D+1},\dots,z_N)
		\nonumber
		\\
		&&
		A^{\{\uz\}}_{p_1 \dots p_D p_{D+1} \dots p_{D+N}}
		=
		\begin{cases}
		A_{q_1 \dots q_D} & p_r = z_r q_r, r \in \{1,\dots,D\} \\
		0 & \text{else}
		\end{cases}
		\label{eq_zp_def2}
\end{eqnarray}
It should be noted that depending on $\uz \in \bbN^N$, an array may be embedded trivially component-wise ($N \leq D$) or even extended dimension-wise ($N > D$). For clarification, consider the examples illustrated in \autoref{fig_emb_1d} for a vector $\ua \in \bbC^{3}$ and in \autoref{fig_emb_2d} for a matrix $\uuA \in \bbC^{(2,3)}$.

\begin{figure}[H]
\centering
\includegraphics[scale=1]{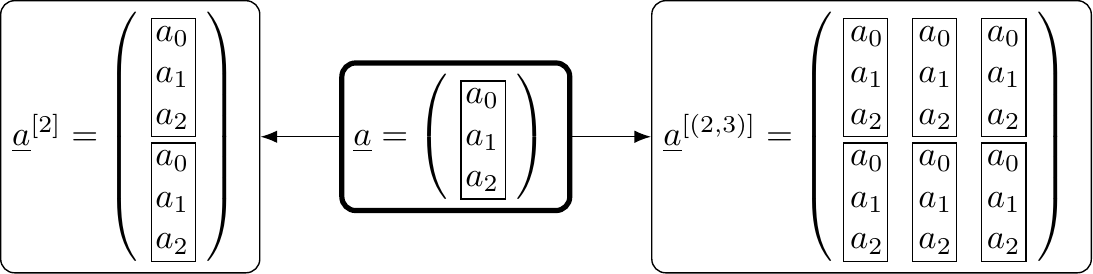}
\caption{Example for the repetition operation of a vector $\ua \in \bbC^3$ (center): $\ua^{[2]} \in \bbC^{6}$ (left); $\ua^{\{(2,3)\}} \in \bbC^{(6,3)}$.}
\label{fig_rep_1d}
\end{figure}

\begin{figure}[H]
\centering
\includegraphics[scale=1]{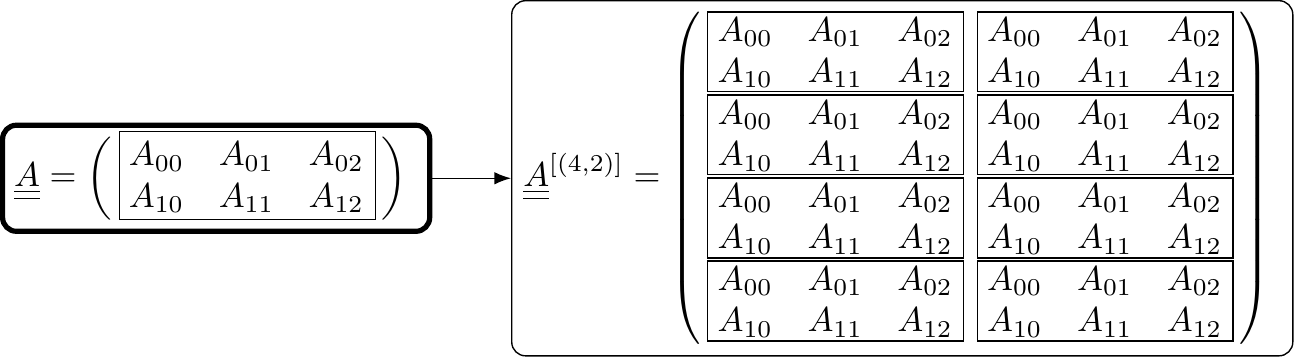}
\caption{Example for the repetition operation of a matrix $\uuA \in \bbC^{(2,3)}$ (left): $\uuA^{[(4,2)]} \in \bbC^{(8,6)}$ (right).}
\label{fig_rep_2d}
\end{figure}

In the following we consider the modulo operator $p|P$ for $P \in \bbN$, which returns the positive remainder on division of $p$ by $P$, e.g., $-2|5 = 3$, $-1|5 = 4$, $0|5 = 0$, $3|5 = 3$ and $7|5 = 2$. We define the repetition operation $\utA^{[z]}$ with $\uz \in \bbN^N$ through its components as
\begin{eqnarray}
	N \leq D : 
		&&
		\utA^{[\uz]} \in \bbC^{\uP'}
		\ , \quad
		\uP' = (z_1P_1,\dots,z_NP_N,P_{N+1},\dots,P_D)
		\nonumber
		\\
		&&
		A^{[\uz]}_{p_1 \dots p_N p_{N+1} \dots p_D}
		=
		A_{(p_1|P_1) \dots (p_N|P_N) p_{N+1} \dots p_D} 
		\label{eq_rep_def1}
		\\
	N > D : 
		&&
		\utA^{[\uz]} \in \bbC^{\uP'}
		\ , \quad
		\uP' = (z_1P_1,\dots,z_DP_D,z_{D+1},\dots,z_N)
		\nonumber
		\\
		&&
		A^{[\uz]}_{p_1 \dots p_D p_{D+1} \dots p_{D+N}}
		=
		A_{(p_1|P_1) \dots (p_D|P_D)}
		\label{eq_rep_def2}
\end{eqnarray}
For clarification, consider the examples illustrated in \autoref{fig_rep_1d} for a vector $\ua \in \bbC^{3}$ and in \autoref{fig_rep_2d} for a matrix $\uuA \in \bbC^{(2,3)}$.

From this point on, we consider any finite dimensional array $\utA \in \bbC^{\uP}$ with $\uP = (P_1,\dots,P_D)$ as the unit cell of a corresponding $(P_1,\dots,P_D)$-periodic array, i.e., we extend the index values to all integers with $A_{p_1 \dots p_D} = A_{(p_1|P_1) \dots (p_D|P_D)}$. For $\utA,\utB \in \bbC^{\uP}$ with $\uP \in \bbN^D$ we consider the circular convolution $\utA * \utB \in \bbC^{\uP}$, defined through its components as
\begin{equation}
	(\utA * \utB)_{p_1 \dots p_D}
	=
	\sum_{q_1 = 0}^{P_1 - 1}
	\dots
	\sum_{q_D = 0}^{P_D - 1}
	A_{q_1 \dots q_D}
	B_{(p_1 - q_1) \dots (p_D - q_D)}
	\ ,
\end{equation}
and the correlation $\utA \circledast \utB \in \bbC^{\uP}$
\begin{equation}
	(\utA \circledast \utB)_{p_1 \dots p_D}
	=
	\sum_{q_1 = 0}^{P_1 - 1}
	\dots
	\sum_{q_D = 0}^{P_D - 1}
	A_{q_1 \dots q_D}
	B_{(q_1 + p_1) \dots (q_D + p_D)}
	\ .
\end{equation}
The multi-dimensional Discrete Fourier Transform (DFT) of an array $\utA$ and the inverse DFT are defined as
\begin{eqnarray}
	\hat\utA = \cF(\utA)
	& , &
	\hat{A}_{p_1 \dots p_D} 
	= 
	\sum_{q_1 = 0}^{P_1 - 1}
	\dots
	\sum_{q_D = 0}^{P_D - 1}
	\exp\left(-i2\pi \left(\frac{p_1q_1}{P_1} + \dots + \frac{p_Dq_D}{P_D}\right)\right)
	A_{q_1 \dots q_D}
	\ ,
	\\
	\utA = \cF^{-1}(\hat{\utA})
	& , &
	A_{p_1 \dots p_D} 
	= 
	\sum_{q_1 = 0}^{P_1 - 1}
	\dots
	\sum_{q_D = 0}^{P_D - 1}
	\frac{
	1}
	{P_1 \dots P_D}
	\exp\left(i2\pi \left(\frac{p_1q_1}{P_1} + \dots + \frac{p_Dq_D}{P_D}\right)\right)
	\hat{A}_{q_1 \dots q_D}
	\ .
\end{eqnarray}
Based on the DFT, the following identities can be shown for $\utA,\utB \in \bbC^{\uP}$ with $\uP \in \bbN^D$, $\uz \in \bbN^D$ and $D \geq 1$
\begin{eqnarray}
	\cF(\utA * \utB) &=& \cF(\utA) \odot \cF(\utB)
	\label{dft_prop_conv}
	\ , \\
	\cF(\utA \circledast \utB) &=& \overline{\cF(\overline{\utA})} \odot \cF(\utB)
	\label{dft_prop_corr}
	\ , \\
	\cF(\utA^{\{\uz\}}) &=& (\cF(\utA))^{[\uz]}
	\label{dft_prop_zp}
	\ , \\
	\cF(\utA^{\{\uz\}} \circledast \utB^{\{\uz\}}) &=& (\cF(\utA \circledast \utB))^{[\uz]} 
	\label{dft_prop_zp2}
	\ .
\end{eqnarray}
The identity \eqref{dft_prop_conv} is the well known convolution theorem for periodic arrays, while \eqref{dft_prop_corr} is an immediate implication of it. Identity \eqref{dft_prop_zp} connects the DFT of a trivially embedded array with the repetition of the DFT of the array. An explicit proof of \eqref{dft_prop_zp} is provided in \autoref{app_proofs}. The identity \eqref{dft_prop_zp2} is a simple implication of \eqref{dft_prop_corr} and \eqref{dft_prop_zp}, its proof is provided in \autoref{app_proofs}. For future purposes, \eqref{dft_prop_zp2} is reformulated for $\utA,\utB \in \bbR^{\uP}$ with $\uP \in \bbN^D$, $\uz \in \bbN^D$ and $D \geq 1$ as
\begin{equation}
	\utA^{\{\uz\}} \circledast \utB^{\{\uz\}}
	= (\utA \circledast \utB)^{\{\uz\}}
	\ ,
	\label{prop_zpcorr}
\end{equation}
such that the correlation of the trivially embedded arrays corresponds to the trivially embedded correlation.

\subsection{Description of structures}
\begin{figure}[H]
\centering
\includegraphics[width=0.3\textwidth]{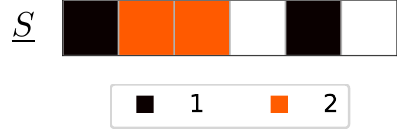}
\caption{Example structure $\uS$ given in \eqref{eq_S_example}}
\label{fig_S_example}
\end{figure}

We consider the array representation of a $D$-dimensional discrete periodic structure comprised of $n$ phases. The indicator $\utI_\alpha \in \bbI^{\uP}$ of phase $\alpha$ denotes the unit cell of the periodic structure with periods $\uP = (P_1,\dots,P_D) \in \bbN^D$. In an $n$-phase structure, the last indicator is uniquely determined through
\begin{equation}
	\utI_n = \utilde{1} - \sum_{\alpha=1}^{n-1} \utI_\alpha
	\ ,
	\label{eq_In}
\end{equation}
where in \eqref{eq_In} $\utilde{1}$ is understood as an array of size $\uP$ with all components equal to 1. Further, we define the structure array $\utS$ as
\begin{equation}
	\utS = \sum_{\alpha = 1}^n \alpha \utI_\alpha
	\ ,
\end{equation}
which can be considered as a position dependent mapping of the phase indices. As an example consider $D=1$, $P=6$, $n=3$ and the instance
\begin{eqnarray}
	\uS &=& (1,2,2,3,1,3) \ , \label{eq_S_example} \\
	\uI_1 &=& (1,0,0,0,1,0) \ , \\
	\uI_2 &=& (0,1,1,0,0,0) \ , \\
	\uI_3 &=& (0,0,0,1,0,1) \ .
\end{eqnarray}
Structures will be visualized in this document through array plots. Hereby, for an $n$-phase structure, the $(n-1)$ first phases will be illustrated through colors and the dependent $n$th phase will be depicted by the white/translucent background. As an example, the structure $\uS$ given in \eqref{eq_S_example} is illustrated in \autoref{fig_S_example}.

\subsection{Two-point correlation}
We define the two-point correlation (2PC) of phases $\alpha_1$ and $\alpha_2$ as
\begin{equation}
	\utC_{\alpha_1\alpha_2} 
		= \utI_{\alpha_1} \circledast \utI_{\alpha_2}
		\in \bbN_0^{\uP}
	\label{eq_2pc}
\end{equation}
In DFT space, the 2PC is equivalently computed based on \eqref{dft_prop_corr} as
\begin{equation}
	\cF(\utC_{\alpha_1\alpha_2})
	= \overline{\cF(\utI_{\alpha_1})}
	\odot \cF(\utI_{\alpha_2})
	\ .
	\label{eq_2pc_dft_alt}
\end{equation}
It should be noted that the array $\utC_{\alpha_1\alpha_2}$ is $\uP$-periodic due to the $\uP$-periodicity of the indicators. Further, it should be noted that the 2PC considered in this work is not normalized by the total number of points, as often done in the literature. The present work considers the non-normalized version of the 2PC given in \eqref{eq_2pc} since it evaluates per definition to exact integers. This simplifies the 2PC-comparison between structures in many programming environments without losing generality. 

In a $n$-phase structure, $n^2$ different 2PC can be computed. The general relation
\begin{equation}
	C_{\alpha_1\alpha_2,p_1 \dots p_D} = C_{\alpha_2\alpha_1,(-p_1) \dots (-p_D)}
	\quad \forall \up \in \bbZ^D
	\label{eq_2pc_gen_rel}
\end{equation}
holds, meaning that, e.g., $\utC_{12}$ is fully determined by $\utC_{21}$. Further, based on \eqref{eq_In}, the 2PC of phase $n$ can be expressed depending on the correlations of the $(n-1)$ phases. For instance, for any $\alpha_1$
\begin{equation}
	\utC_{\alpha_1 n} 
	= \utI_{\alpha_1} \circledast \utI_n
	= \utI_{\alpha_1} \circledast \left(\utilde{1} - \sum_{\alpha_2=1}^{n-1} \utI_{\alpha_2} \right)
	= \utI_{\alpha_1} \circledast \utilde{1} 
		- \sum_{\alpha_2=1}^{n-1} \utC_{\alpha_1 \alpha_2}
	= (C_{\alpha_1 \alpha_1,0...0}) \utilde{1}
		- \sum_{\alpha_2=1}^{n-1} \utC_{\alpha_1 \alpha_2}
	\label{eq_2pc_last}
\end{equation}
holds. It can be concluded that knowledge of the $n(n-1)/2$ independent correlations $\utC_{\alpha_1\alpha_2}$ for $\alpha_1 \in \{1,\dots,n-1\}$ and $\alpha_2 \in \{\alpha_1,\dots,n-1\}$ suffices for the complete determination of all remaining 2PC. It is shortly remarked that in \cite{Niezgoda2008} it is claimed that knowledge of only $(n-1)$ correlations suffices for the unique determination of all 2PC. In \cite{Niezgoda2008} several relations are presented based on the DFT of the 2PC, but no algebraic proof is delivered that the presented relations indeed yield a uniquely solvable system of equations for arbitrary $n$ and structural shapes. All equations of \cite{Niezgoda2008} have been taken into account and, to the best knowledge of the authors of the present work, the claim of \cite{Niezgoda2008} is not true, in general. A discussion and explicit counterexamples illustrating the issues of \cite{Niezgoda2008} are provided in \autoref{app_niezgoda_counterexample}. It is then concluded that the determination of the minimal number of independent 2PCs still requires more investigation. This is important since the work of \cite{Niezgoda2008} has been relevant to several investigations, see, e.g., \cite{Panchal2013a}, \cite{Miyazawa2019} and \cite{Ashton2020}. In the following, the comparison of all $(n-1)n/2$ independent 2PCs is used to provide unambiguous results. 

Consider two structures $\utS(1)$ and $\utS(2)$ with corresponding indicators $\utI_\alpha(s)$ for $s \in \{1,2\}$. Two structures $\utS(1)$ and $\utS(2)$ are referred to as 2PC-equivalent, shortly denoted as 
\begin{equation}
	\utS(1) \sim \utS(2) 
	\ ,
	\label{eq_2pceq_def}
\end{equation}
if all 2PC $\utC_{\alpha_1\alpha_2}(1)$ corresponding to $\utS(1)$ and its indicators $\utI_\alpha(1)$ and all $\utC_{\alpha_1\alpha_2}(2)$ corresponding to $\utS(2)$ are identical, i.e., $\utC_{\alpha_1\alpha_2}(1) = \utC_{\alpha_1\alpha_2}(2) \ \forall \alpha_1,\alpha_2 \in \{1,\dots,n\}$. Again, due to the evaluation of the non-normalized version of the 2PC to exact integers, 2PC-equivalence of given structures can easily be tested in many programming languages without any ambiguities. The equality of $n(n-1)/2$ independent 2PCs of two structures suffices for the 2PC-equivalence of the structures.

\subsection{Higher-order correlations}

Naturally, correlations between multiple points can also be evaluated for structures. The general $M$-point correlation $\utC_{\langle M \rangle \ualpha}$ ($M$PC) for $M \geq 2$ for a $D$-dimensional periodic structure $\utS$ with periods vector $\uP = (P_1,\dots,P_D) \in \bbN^D$, phase vector $\ualpha = (\alpha_1,\dots,\alpha_M) \in \bbN^M$ and indicators $\utI_{\alpha_j}, \alpha_j \in \ualpha$ is defined component-wise as
\begin{align}
	(\utC_{\langle M \rangle \ualpha})_{\up_1 \dots \up_{M-1}}
	&=
	\sum_{q_1 = 0}^{P_1 - 1}
	\dots
	\sum_{q_D = 0}^{P_D - 1}
	(\utI_{\alpha_1})_{\uq}
	(\utI_{\alpha_2})_{\uq + \up_1}
	\dots
	(\utI_{\alpha_n})_{\uq + \up_{M-1}}
	\ .
	\label{eq_mpc}
\end{align}
For example, for $D = 3$, the 3PC of the phases $\ualpha = (2,3,2)$ is a six-dimensional array $\utC_{\langle 3 \rangle 232} \in \bbN_0^{\uP}$, $\uP = (P_1,P_2,P_3)$ with components $(\utC_{\langle 3 \rangle 232})_{\up_1 \up_2}$. More explicitly, defining $\up_1 = (p_{11},p_{12},p_{13})$ and $\up_2 = (p_{21},p_{22},p_{23})$, the components are computed as follows
\begin{equation}
	(\utC_{\langle 3 \rangle 232})_{p_{11}p_{12}p_{13}p_{21}p_{22}p_{23}}
	=
	\sum_{q_1 = 0}^{P_1 - 1}
	\sum_{q_2 = 0}^{P_2 - 1}
	\sum_{q_3 = 0}^{P_3 - 1}
	(\utI_2)_{q_1 q_2 q_3}
	(\utI_3)_{(q_1 + p_{11}) (q_1 + p_{12}) (q_3 + p_{13})}
	(\utI_2)_{(q_1 + p_{21}) (q_1 + p_{22}) (q_3 + p_{23})}
	.
\end{equation}
The definition \eqref{eq_mpc} is in accordance with the definition of the 2PC given in \eqref{eq_2pc}, i.e, $\utC_{\langle 2 \rangle \alpha_1 \alpha_2} = \utC_{\alpha_1 \alpha_2}$. In this work, the main focus in on the 2PC, while the $M$PC will be evaluated only for few, specific examples. Note that the number of independent components of the $M$PC is proportional to $\norm{\uP}^{M-1} n^M$, i.e., it grows exponentially with $M$.

\subsection{Inheritance of 2PC-equivalence}
\label{sec_inheritance}

\paragraph{Starting point.} We assume that two bi-phasic 2PC-equivalent structures $\utS(1) \sim \utS(2)$ are given, i.e.,
\begin{equation}
	\utC_{11}(1) = \utC_{11}(2)
	\ , \quad
	\utC_{12}(1) = \utC_{12}(2)
	\ , \quad
	\utC_{21}(1) = \utC_{21}(2)
	\ , \quad
	\utC_{22}(1) = \utC_{22}(2)
	\label{eq_inh_2ph_C}
\end{equation}
hold. The structures contain only two phases, i.e., it suffices, in principle, to  have access to $\utI_1(s)$ for each structure $s \in \{1,2\}$. Due to the dependencies of the 2PC, $\utC_{11}(1) = \utC_{11}(2)$ suffices for the 2PC-equivalence. 

\paragraph{Phase extension through trivial embedding.} Consider a structure $\utS \in \{\utS(1),\utS(2)\}$ and indicators indicators $\utI_1$ and $\utI_2$ corresponding to the chosen $\utS$. We define new indicators $\utI'_\alpha$ base on the trivially embedded indicators of the structure $\utS$ for phases $\alpha \in \{1,2\}$
\begin{equation}
	\utI'_\alpha
	= \utI_\alpha^{\{\uz\}}
	\ , \quad
	\alpha \in \{1,2\}
	\ .
\end{equation}
Next, we define the additional new phase indicator as
\begin{equation}
	\utI'_3 = 1 - \sum_{\alpha=1}^2 \utI'_\alpha
\end{equation}
such that we obtain a new three-phase structure
\begin{equation}
	\utS' = \sum_{\alpha=1}^3 \alpha \utI'_\alpha
	\ .
\end{equation}
Naturally, the phase-extended structure $\utS'$ corresponds to a simple trivial embedding of the original structure and inserting the number 3 in the positions with 0. The 2PCs of the new structure $\utS'$ are related to the old ones for $\alpha_1,\alpha_2 \in \{1,2\}$ due to the property \eqref{prop_zpcorr} as follows
\begin{equation}
	\utC'_{\alpha_1\alpha_2}
	= \utI'_{\alpha_1} \circledast \utI'_{\alpha_2}
	= \utI^{\{\uz\}}_{\alpha_1} \circledast \utI^{\{\uz\}}_{\alpha_2}
	= (\utC_{\alpha_1\alpha_2})^{\{\uz\}}
	\ .
	\label{eq_phe_corr}
\end{equation}
The relation \eqref{eq_phe_corr} makes then clear that if $\utS(1)$ is phase-extended and denoted as $\utS'(1)$, and $\utS(2)$ is phase-extended and denoted as $\utS'(2)$, then all corresponding 2PC for $\alpha_1,\alpha_2 \in \{1,2\}$ fulfill $\utC'_{\alpha_1\alpha_2}(1) = \utC'_{\alpha_1\alpha_2}(2)$. More explicitly, for $n=3$ in the new structures $\utS'(1)$ and $\utS'(2)$, we have 
\begin{equation}
	\utC'_{11}(1) = \utC'_{11}(2)
	\ , \quad
	\utC'_{12}(1) = \utC'_{12}(2)
	\ , \quad
	\utC'_{22}(1) = \utC'_{22}(2)
\end{equation}
for the $(n-1)n/2 = 3$ independent 2PC. This is, as discussed for the definition \eqref{eq_2pceq_def} based on \eqref{eq_2pc_gen_rel} and \eqref{eq_2pc_last}, implies 
\begin{equation}
	\utC'_{\alpha_1\alpha_2}(1) = \utC'_{\alpha_1\alpha_2}(2) 
	\quad \forall \alpha_1,\alpha_2 \in \{1,2,3\}
\end{equation}
and is, therefore, sufficient for the 2PC-equivalence of the new three-phase structures structures, i.e., 
\begin{equation}
	\utS(1) \sim \utS(2)
	\quad \Rightarrow \quad
	\utS'(1) \sim \utS'(2)
\end{equation}
holds. In terms of graph theory, the structures $\utS(1) \sim \utS(2)$ are regarded as \emph{parent} structures, while the derived ones $\utS'(1) \sim \utS'(2)$ are denoted as \emph{child} structures inheriting the 2PC-equivalence. It should be noted that the just described reasoning for the extension from 2 to 3 phases can be repeated for $3 \to 4$, $4 \to 5$ and so on. This means that
\begin{itemize}
\item[(i)] a phase-extension of 2PC-equivalent $n$-phase structures through trivial embedding yields 2PC-equivalent $(n+1)$-phase structures, 
\item[(ii)] 2PC-equivalent structures with arbitrary number of phases can always be generated starting from two-phase 2PC-equivalent structures, and
\item[(iii)] the described phase extension goes along with an increase of the size of the structure, possibly with increasing spatial dimension  $D$.
\end{itemize}
For example, consider $\uI_1 \in \bbI^{12}$ and $\uI^{\{(2,4,5)\}}_1 \in \bbI^{(24,4,5)}$, where a one-dimensional two-phase structure can be used in order to generate a three-dimensional three-phase structure. Corresponding one-dimensional 2PC-equivalent structures would then generate three-dimensional 2PC-equivalent structures.

\paragraph{Kernel-based extension.} For a $n$-phase structure $\utS$ with corresponding indicators $\utI_\alpha$ we assume that a kernel list $\rmK = \{\utK_1,\dots,\utK_n\}$ with $\utK_\alpha \in \bbI^{\uz} \ \forall \alpha \in \{1,\dots,n\}$ is given. Based on the dimensions $\uz \in \bbN^N$ of the given kernels, we extend the structure through trivial embedding of the indicators $\utI_\alpha^{\{\uz\}} \in \bbI^{\uP'}$ and extend the kernels through appended trivial embedding $\utK_\alpha^0 \in \bbI^{\uP'}$ to match the new structure dimensions as follows
\begin{eqnarray}
	N \leq D:
	&&
	\uP' = (z_1P_1,\dots,z_NP_N,P_{N+1},\dots,P_D)
	\ ,
	\nonumber
	\\
	&&
	\utK^0_{\alpha,p'_1 \dots p'_N p'_{N+1} \dots p'_D}
	= \begin{cases}
	K_{\alpha,p'_1 \dots p'_N} & p'_r \in \{0,\dots,z_r-1\} \ \forall r \in \{1,\dots,N\}
	\\
	& \text{and} \ p'_t = 0 \ \forall t \in {N+1,\dots,D}
	\\
	0 & \text{else}
	\end{cases}
	\ , 
	\\
	N > D:
	&&
	\uP' = (z_1P_1,\dots,z_DP_D,z_{D+1},\dots,z_N)
	\ ,
	\nonumber
	\\
	&&
	\utK^0_{\alpha,p'_1 \dots p'_N}
	= \begin{cases}
	K_{\alpha,p'_1 \dots p'_N} & p'_r \in \{0,\dots,z_r-1\} \ \forall r \in \{1,\dots,N\}
	\\
	0 & \text{else}
	\end{cases}
	\ .
\end{eqnarray}
Based on the dimension $\uz$ of the kernels and corresponding $\utK^0_{\alpha}$, the $(n+1)$ indicators
\begin{equation}
	\utI'_\alpha = \utK^0_\alpha * \utI_\alpha^{\{\uz\}} \in \bbI^{\uP'}
	, \
	\alpha \in \{1,\dots,n\}
	\quad , \quad
	\utI'_{n+1} = \utilde{1} - \sum_{\alpha=1}^n \utI'_\alpha
\end{equation}
describe the kernel-based extended $(n+1)$-phase structure $\utS' \in \bbN^{\uP'}$. The corresponding 2PC of the new structure $\utS'$ are related to the original ones due to \eqref{eq_2pc_dft_alt}, \eqref{dft_prop_conv} and \eqref{dft_prop_zp} as follows
\begin{eqnarray}
	\cF(\utC'_{\alpha_1 \alpha_2})
	&=& \overline{\cF(\utI'_{\alpha_1})} \odot \cF(\utI'_{\alpha_2})
	\nonumber
	\\
	&=&
	\overline{\cF\left(\utK^0_{\alpha_1} * \utI_{\alpha_1}^{\{\uz\}}\right)}
	\odot \cF\left(\utK^0_{\alpha_2} * \utI_{\alpha_2}^{\{\uz\}}\right)
	\nonumber
	\\
	&=& 
	\overline{\cF(\utK^0_{\alpha_1}) \odot (\cF(\utI_{\alpha_1}))^{[\uz]}}
	\odot \cF(\utK^0_{\alpha_2}) \odot (\cF(\utI_{\alpha_2}))^{[\uz]}
	\nonumber
	\\
	&=&
	\overline{\cF(\utK^0_{\alpha_1})}
	\odot \cF(\utK^0_{\alpha_2}) 
	\odot \overline{(\cF(\utI_{\alpha_1}))^{[\uz]}}
	\odot (\cF(\utI_{\alpha_2}))^{[\uz]}	
	\nonumber
	\\
	&=& \overline{\cF(\utK^0_{\alpha_1})} \odot \cF(\utK^0_{\alpha_2})
	\odot (\cF(\utC_{\alpha_1 \alpha_2}))^{[\uz]}
	\ .
	\label{eq_ker_corr}
\end{eqnarray}
Starting from the $n$-phase parent structures $\utS(1) \sim \utS(2)$ with $\utC_{\alpha_1 \alpha_2} = \utC_{\alpha_1 \alpha_2} \ \forall \alpha_1,\alpha_2 \in \{1,\dots,n\}$, the 2PC of the structures $\utS'(1) \sim \utS'(2)$ fulfill due to \eqref{eq_ker_corr}
\begin{equation}
	\utC'_{\alpha_1 \alpha_2}
	= \utC'_{\alpha_1 \alpha_2}
	\quad \forall \alpha_1,\alpha_2 \in \{1,\dots,n\}
\end{equation}
which is sufficient for $\utC'_{\alpha_1 \alpha_2} = \utC'_{\alpha_1 \alpha_2} \ \forall \alpha_1,\alpha_2 \in \{1,\dots,n+1\}$. Therefore, for parent structures $\utS(1) \sim \utS(2)$, \eqref{eq_ker_corr} implies that the kernel-extended child structures $\utS'(1)$ and $\utS'(2)$ inherit the 2PC-equivalence, i.e., $\utS'(1) \sim \utS'(2)$ holds. Kernels with components $K_{\alpha,p_1\dots p_N} = 1$ simply return the original structure in a higher resolution with possible re-scaling, i.e., a change of aspect ratio of the axes. 
It can therefore be concluded that
\begin{itemize}
\item[(i)] 2PC-equivalence is invariant under appropriate application of kernels, and
\item[(ii)] 2PC-equivalence is invariant under arbitrary discrete rescaling.
\end{itemize}

\paragraph{Phase coalescence.} Is is shortly remarked that 
\begin{itemize}
\item[(i)] any subset of the phases within 2PC equivalent structures can be coalesced to a single phase, without affecting the 2PC-equivalence.
\end{itemize}
For instance, consider a given four-phase structure $\utS$ with phase indicators $\utI_\alpha, \alpha \in \{1,2,3,4\}$. One option would be to coalesce the three first phases and generate a new two-phase structure with indicators $\utI'_1 = \sum_{\alpha=1}^3 \utI_\alpha$ and $\utI'_2 = \utI_4$. The 2PC of phases $(\alpha_1,\alpha_2) = (1,2)$ of the new two-phase structure is computed as
\begin{equation}
	\utC'_{12} 
	= \utI'_1 \circledast \utI'_2
	= (\utI_1 + \utI_2 + \utI_3) \circledast \utI_4
	= \utC_{14} + \utC_{24} + \utC_{34}
	\ .
	\label{eq_phc_c12}
\end{equation}
For two 2PC-equivalent structures, \eqref{eq_phc_c12} makes clear that not only the 2PC for $(\alpha_1,\alpha_2) = (1,2)$ of the new two-phase structures are then equal, but all of the 2PC, such that the new two-phase structures are also 2PC-equivalent. This small example immediately makes clear that for given $\utS(1) \sim \utS(2)$ any choice of phase coalescence inducing new corresponding structures $\utS'(1)$ and $\utS'(2)$ preserves the 2PC-equivalence, i.e., $\utS'(1) \sim \utS'(2)$ holds for arbitrary phase coalescence.

\section{Examples and application}
\label{sec_examples_applications}

\subsection{Root structures}

The generation of 2PC-equivalent structures based on the inheritance properties proven in \autoref{sec_inheritance} (phase extension, kernel-based extension and phase coalescence) requires given 2PC-equivalent structures. Small 2PC-equivalent structures with a low number of phases can easily be searched for by brute force. These small structures can be used in order to generate derived 2PC-equivalent structures. Hereby, for given dimension vector $\uP \in \bbN^D$ for $D \in \{1,2,3\}$ and number of phases $n \leq 3$ we search for 2PC-equivalent structures as described through the following steps:
\begin{enumerate}
\item[(S.1)] Based on $\uP$ and $n$ generate all possible $n_\rmS$ structures $\cS = \{\utS(1),\dots,\utS(n_\rmS)\}$.
\item[(S.2)] Denote $\utC(s)$ the array containing all $n(n-1)/2$ independent 2PC corresponding to the $s$th structure $\uS(s)$. Compute all 2PC and form the corresponding set $\cC = \{\utC(1),\dots,\utC(n_\rmS)\}$.
\item[(S.3)] Set $s_{\mathrm{ref}}=1$, i.e., select the first structure as start reference
\item[(S.4)] Extract the index set of candidates $\Sigma$ containing all 2PC matching $\utC(s_{\mathrm{ref}})$, i.e., $\Sigma = \{s \in \{s_{\mathrm{ref}} + 1,\dots,n_\rmS\} : \utC(s_{\mathrm{ref}}) = \utC(s)\}$.
\item[(S.5)] Denote by $\mathrm{unrel}(\utS(s),\utS(s'))$ a boolean operation returning 1 if the structures $\utS(s)$ and $\utS(s')$ are \emph{not} related (i.e., unrelated) by any periodic shift, any axis reflection, any phase interchange or any combination of these operations, and 0 otherwise. Reduce the candidate set $\Sigma$ to those unrelated to $\utS(s_{\mathrm{ref}})$, i.e., define $\Sigma^* = \{s \in \Sigma : \mathrm{unrel}(\utS(s),\utS(s_\mathrm{ref})) = 1\}$.
\item[(S.6)] If $\Sigma^*$ is nonempty, extract the largest set of pair-wise unrelated structures $\Sigma^{**} \subset \Sigma^*$ (i.e., $\mathrm{unrel}(\utS(s),\utS(s')) = 1 \ \forall s,s' \in \Sigma^{**}, s \neq s'$) and return the 2PC-equivalent structure set $\cS^{**} = \{\utS(s_{\mathrm{ref}})\} \cup \{\utS \in \cS : \utS(s) \ , \ s \in \Sigma^{**}\}$. Otherwise, if $s_{\mathrm{ref}} + 1 \leq n_\mathrm{S}$ increase $s_{\mathrm{ref}} = s_{\mathrm{ref}} + 1$ and go to (S.4), else terminate search (no 2PC-equivalent structures exist for the given $\uP$ and $n$).
\end{enumerate}
Structures found based on the just described procedure will be referred to as root structures from this point on.

\subsection{Two-phase and $n$-phase structures}

\begin{figure}[H]
\centering
\includegraphics[width=0.9\textwidth]{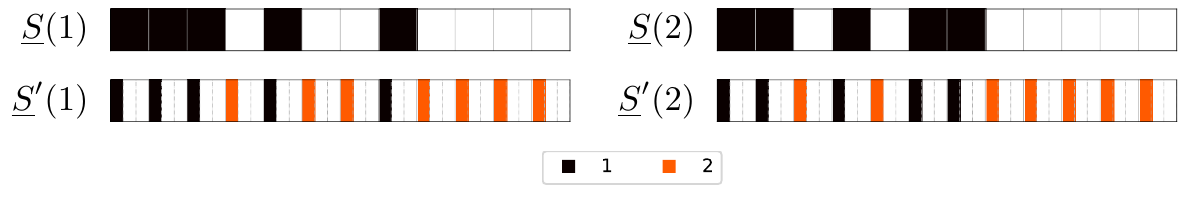}
\caption{
Root structures $\uS(1) \sim \uS(2)$ and corresponding phase extended structures $\uS'(1) \sim \uS'(2)$ based on $\uz = 3$.
}
\label{fig_ex_1d}
\end{figure}

\paragraph{One-dimensional example.}
For the period $P=12$ and number of phases $n=2$, consider the given 2PC-equivalent root vectors/structures
\begin{equation}
	\uS(1)
	=
	(1,1,1,2,1,2,2,1,2,2,2,2)
    \ , \quad
    \uS(2)
	=
	(1,1,2,1,2,1,1,2,2,2,2,2)
    \ .
    \label{eq_1d_Isum12}
\end{equation}
The root structures $\uS(1),\uS(2) \in \bbN^{12}$ are displayed in \autoref{fig_ex_1d}, together with the corresponding phase extended three-phase structures based on the trivial embedding $\uz = 3$ (i.e., $\uS'(1), \uS'(2) \in \bbN^{36}$).

\paragraph{Two-dimensional example.}
Consider the given 2PC-equivalent root matrices/structures for $\uP = (P_1,P_2) = (4,3)$ and $n=2$
\begin{equation}
	\uuS(1)
	=
	\begin{pmatrix}
	1 & 2 & 2 \\
    2 & 2 & 1 \\
    2 & 1 & 1 \\
    2 & 2 & 1
    \end{pmatrix}
    \ , \quad
    \uuS(2)
	=
	\begin{pmatrix}
	1 & 2 & 2 \\
    2 & 1 & 1 \\
    1 & 2 & 2 \\
    1 & 2 & 2
    \end{pmatrix}
    \ .
    \label{eq_2d_Isum12}
\end{equation}

\begin{figure}[H]
\centering
\includegraphics[width=0.98\textwidth]{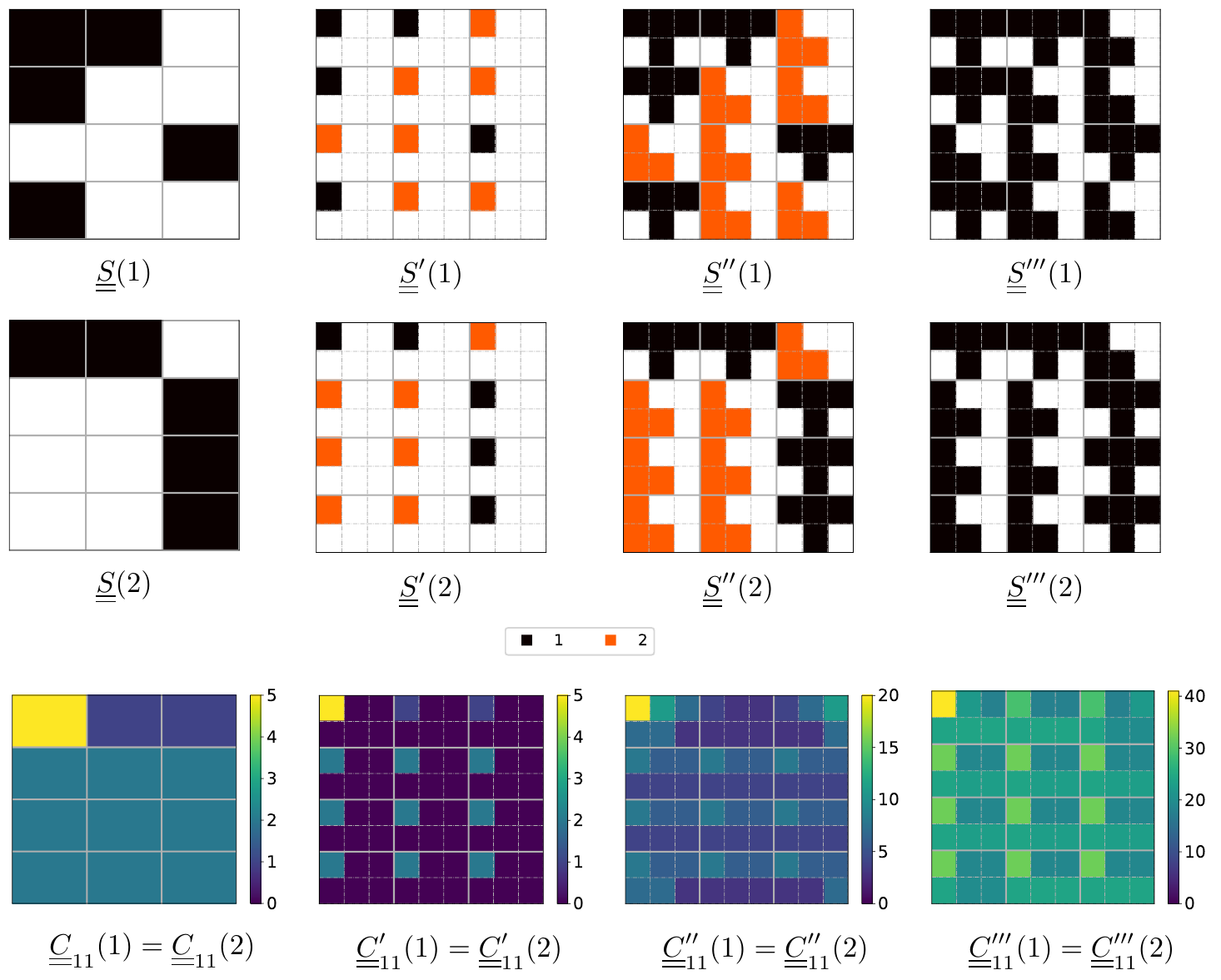}
\caption{
Root structures $\uuS(1) \sim \uuS(2)$%
, trivial embedding extended structures $\uuS'(1) \sim \uuS'(2)$%
, kernel-based extended structures $\uuS''(1) \sim \uuS''(2)$%
, and phase coalesced structures $\uuS'''(1) \sim \uuS'''(2)$ by coalescence of phases 1 and 2 of $\uuS''(1) \sim \uuS''(2)$; The 2PC for $(\alpha_1,\alpha_2) = (1,1)$ of the corresponding structures is displayed at the bottom of the corresponding column.
}
\label{fig_2d_example}
\end{figure}

The two-phase root structures $\uuS(1) \sim \uuS(2) \in \bbN^{(4,3)}$ given by \eqref{eq_2d_Isum12} are illustrated in \autoref{fig_2d_example}. A phase-extension of the two-phase structures to three phases based on a trivial embedding with $\uz = (2,3)$ is shown in \autoref{fig_2d_example} by $\uuS'(1) \sim \uuS'(2) \in \bbN^{(8,9)}$. It should be noted that the axis ratio is changed from 4:3 to 8:9, i.e., a relative stretching in the second axis is performed. The property \eqref{prop_zpcorr} and its implication \eqref{eq_phe_corr} can be seen in the 2PC $\uuC_{11}$ and $\uuC'_{11}$ of the root and phase extended structures at the bottom of the corresponding columns in \autoref{fig_2d_example}. Application of the kernels
\begin{equation}
	\uuK_1 
	=
	\begin{pmatrix}
	1 & 1 & 1 \\
	0 & 1 & 0 
	\end{pmatrix}
	\ , \quad
	\uuK_2 
	=
	\begin{pmatrix}
	1 & 0 & 0 \\
	1 & 1 & 0 
	\end{pmatrix}
	\ ,
\end{equation}
on the embedded phases 1 and 2, respectively, yields the 2PC-equivalent structures $\uuS''(1) \sim \uuS''(2)$ shown in \autoref{fig_2d_example}. It should be noted that in case of $K_{1,p_1p_2} = 1$ and $K_{2,p_1p_2} = 1$, the root two-phase structures $\uuS(1) \sim \uuS(2)$ would have been returned, only in a higher resolution and with some relative stretching along the second dimension. This emphasizes that from given 2PC-equivalent structures, derived child structures with arbitrary resolution and size changes are also 2PC-equivalent. Finally, $\uuS'''(1) \sim \uuS'''(2)$ in \autoref{fig_2d_example} show a phase coalescence of phases 1 and 2 of the structures $\uuS''(1) \sim \uuS''(2)$, which naturally drastically changes the phase volume fractions and topology of the structures, but the 2PC-equivalence is still preserved.

\begin{figure}[h]
\centering
\begin{tabular}{ccc}
\includegraphics[scale=0.4]{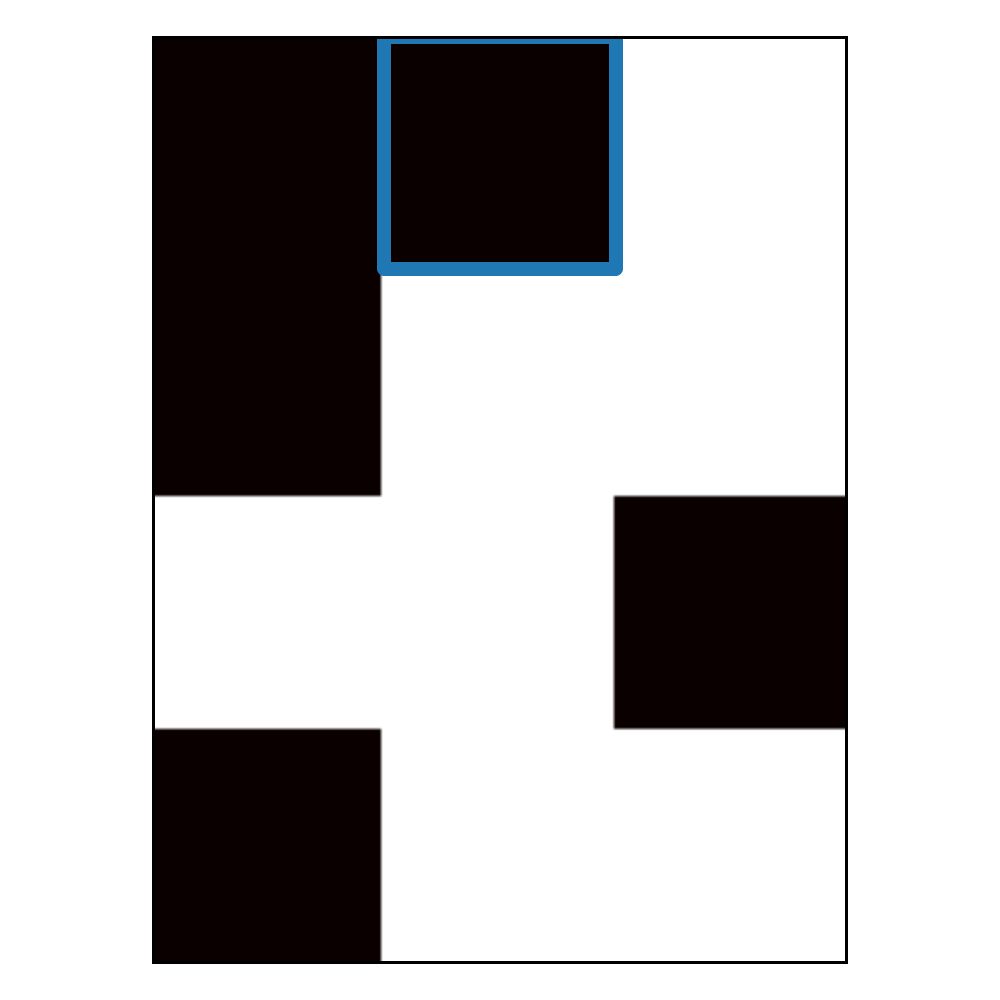}
&
\includegraphics[scale=0.4]{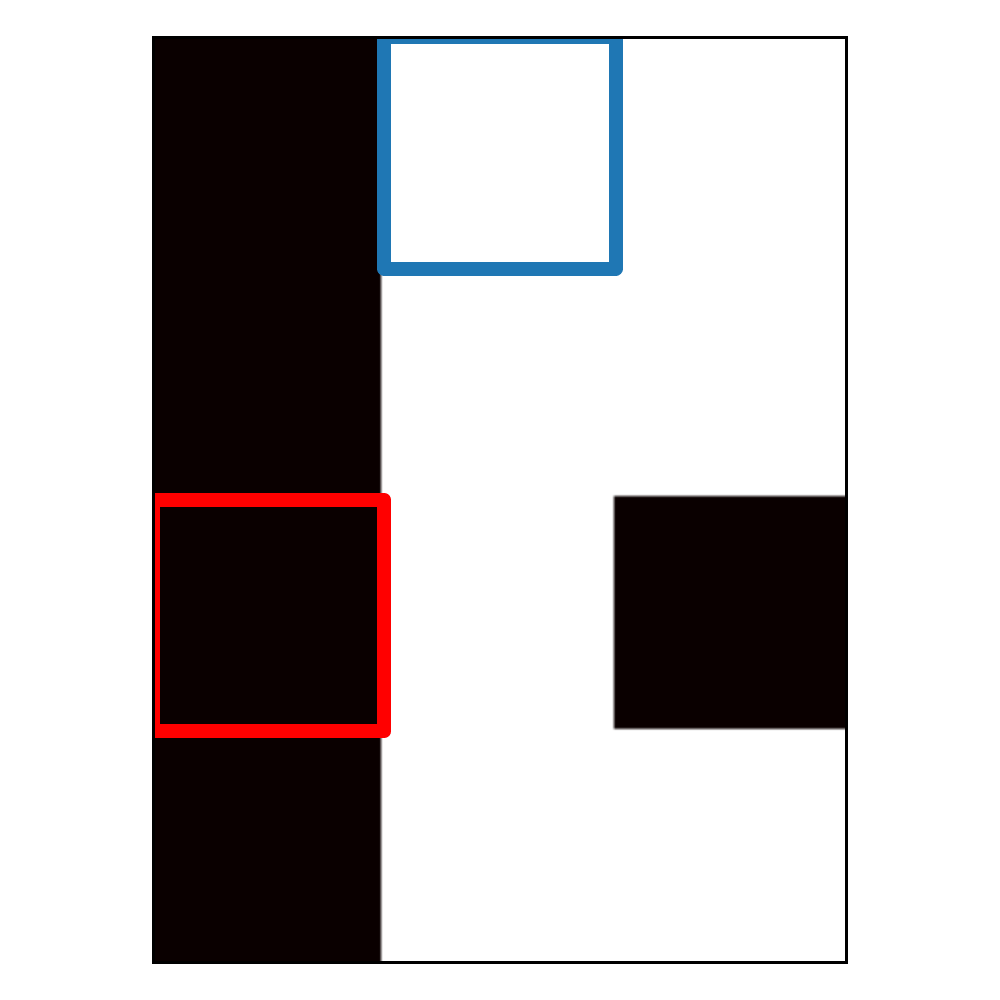}
&
\includegraphics[scale=0.4]{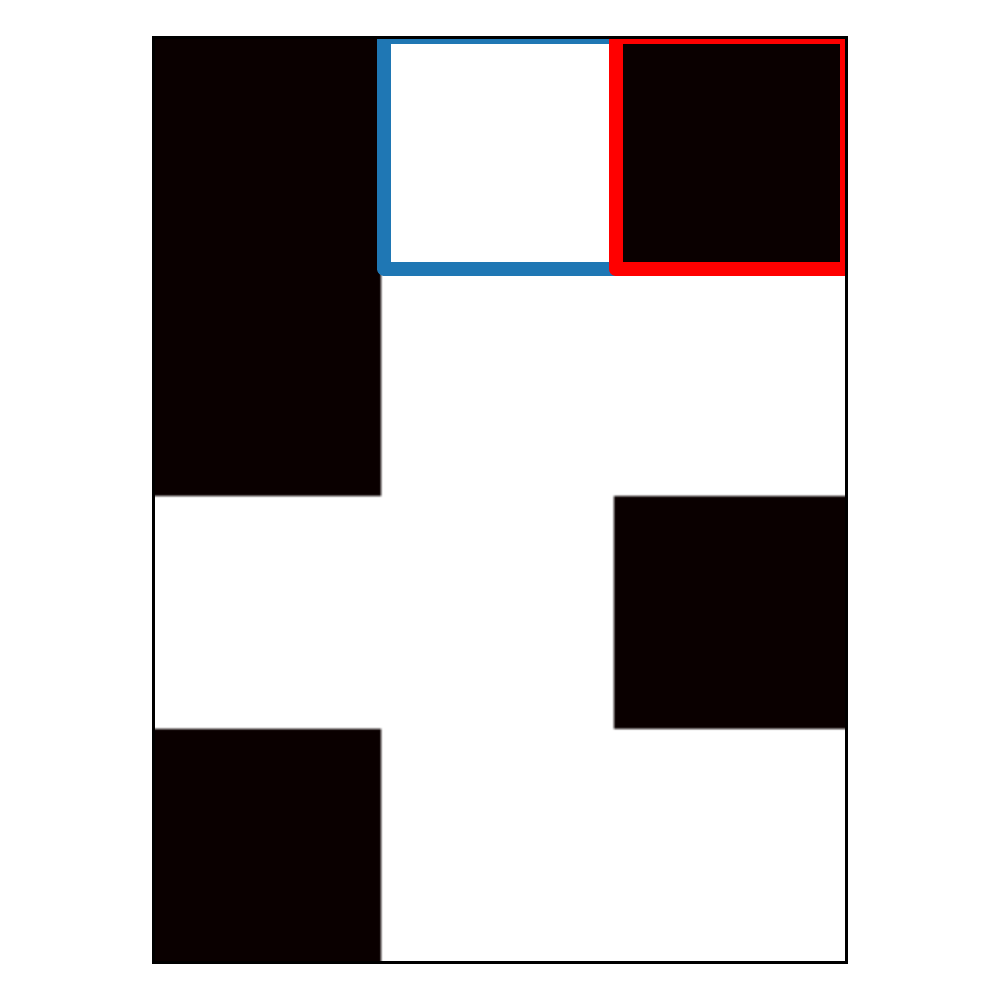}
\\
$\uuS(1)$
&
$\uuS^\mathrm{f}(1)$
&
$\uuS^\mathrm{f}(2)$
\end{tabular}
\caption{Single pixel flips of the structure $\uuS(1)$ yielding non-2PC-equivalent structures $\uuS^\mathrm{f}(1)$ (percolating) and $\uuS^\mathrm{f}(2)$.}
\label{fig_flipping}
\end{figure}

It is interesting to visually note that the structures $\uuS(1) \in \bbN^{(4,3)}$ and $\uuS(2) \in \bbN^{(4,3)}$ displayed in \autoref{fig_2d_example} differ by a single pixel flip, i.e., if the pixel at positions $(p_1,p_2) = (0,1)$ and $(p'_1,p'_2) = (2,1)$ (keep in mind $p_1,p'_1 \in \{0,1,2,3\}$ and $p_2,p'_2 \in \{0,1,2\}$ hold for the current structures) are flipped in $\uuS(1)$, then $\uuS(2)$ is obtained (with the corresponding periodic translation). It can, therefore, be stated that single pixel flips may exist which conserve the 2PC. However, not every single pixel flip preserves the 2PC-equivalence and, more importantly, some single pixel flips may cause important changes in the structures. For example, as shown in \autoref{fig_flipping}, flipping the positions $(p_1,p_2) = (0,1)$ (marked in blue in \autoref{fig_flipping}) and $(p'_1,p'_2) = (3,0)$ in $\uuS(1)$ yields the structure $\uuS^\mathrm{f}(1)$, which percolates along its first column. Percolation in structures is an important feature which can strongly affect the properties of physical systems. This will be demonstrated in the subsequent section. While some pixel flips may drastically change the topology of the structure, as in $\uuS^\mathrm{f}(1)$, others may induce negligible changes. The structure $\uuS^\mathrm{f}(2)$ illustrated in \autoref{fig_flipping} corresponds to $\uuS(1)$ by flipping $(p_1,p_2) = (0,1)$ and $(p'_1,p'_2) = (0,2)$ and, from a visual point of view, its influence on the topology changes may be considered as negligible with respect to the effective properties of the physical structure. This will be discussed in the following section.

For higher number of structure elements, such as for $\uuS'''(1)$ and $\uuS'''(2)$, displayed in \autoref{fig_2d_example}, single pixel flips conserving the 2PC are unknown to the authors. But due to the connection to the original structures $\uuS(1)$ and $\uuS(2)$ it becomes evident that a "master"-pixel flip corresponding to the single pixel flip for $\uuS(1)$ and $\uuS(2)$ but with the corresponding pixels in $\uuS'''(1)$ and $\uuS'''(2)$ preserves the 2PC.

\begin{figure}[H]
\centering
\includegraphics[scale=0.4]{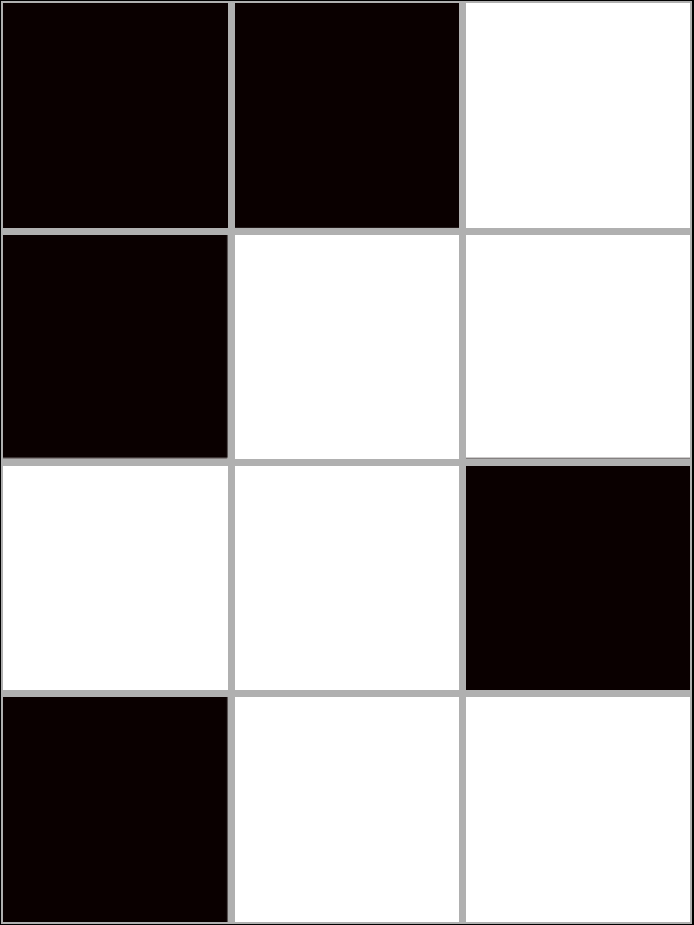} \quad
\includegraphics[scale=0.4]{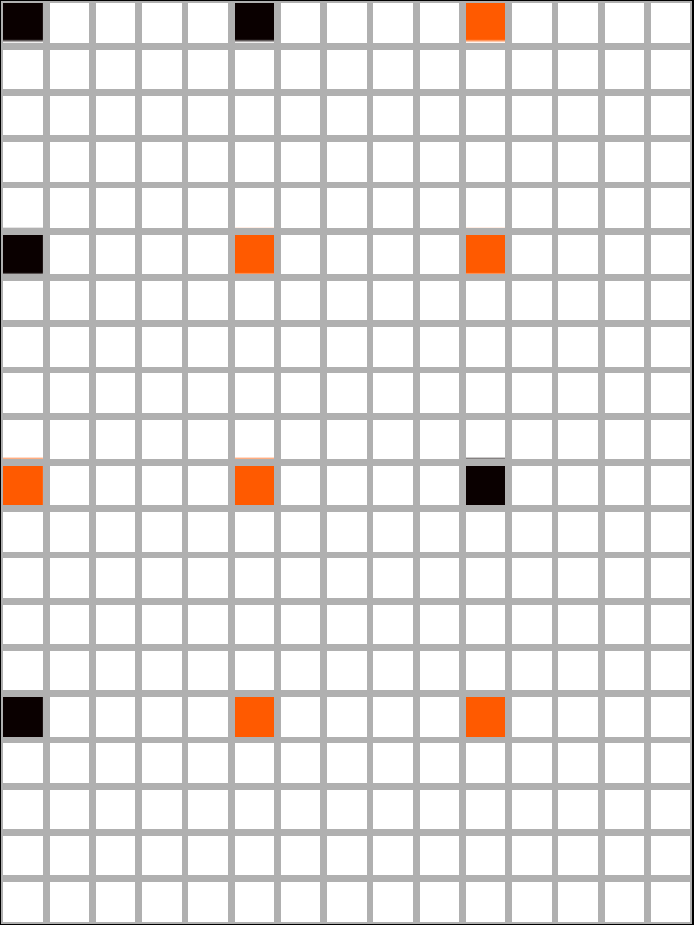} \quad
\includegraphics[scale=0.4]{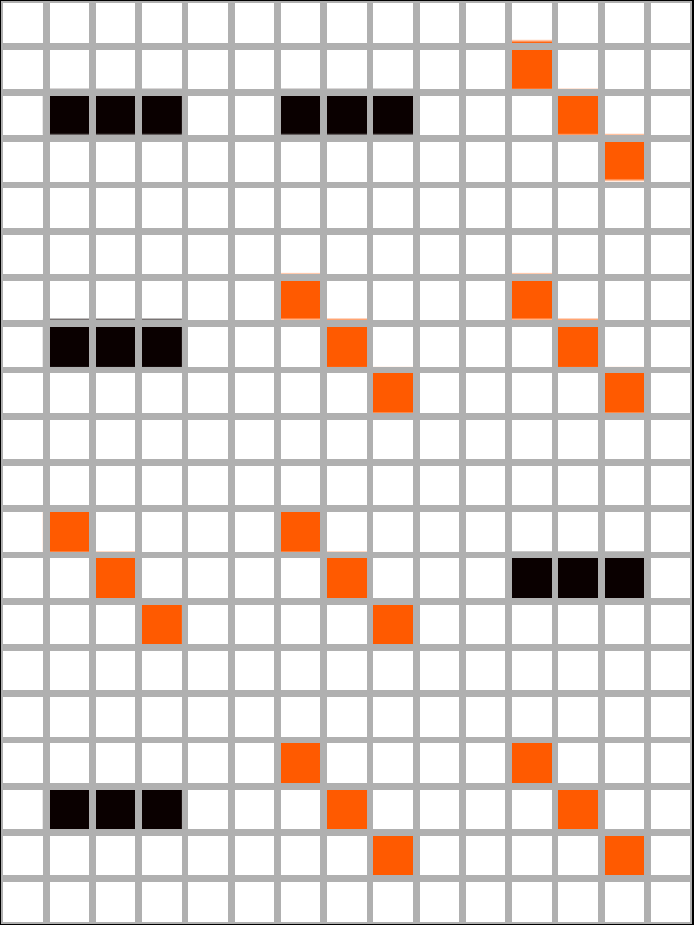} \quad
\includegraphics[scale=0.4]{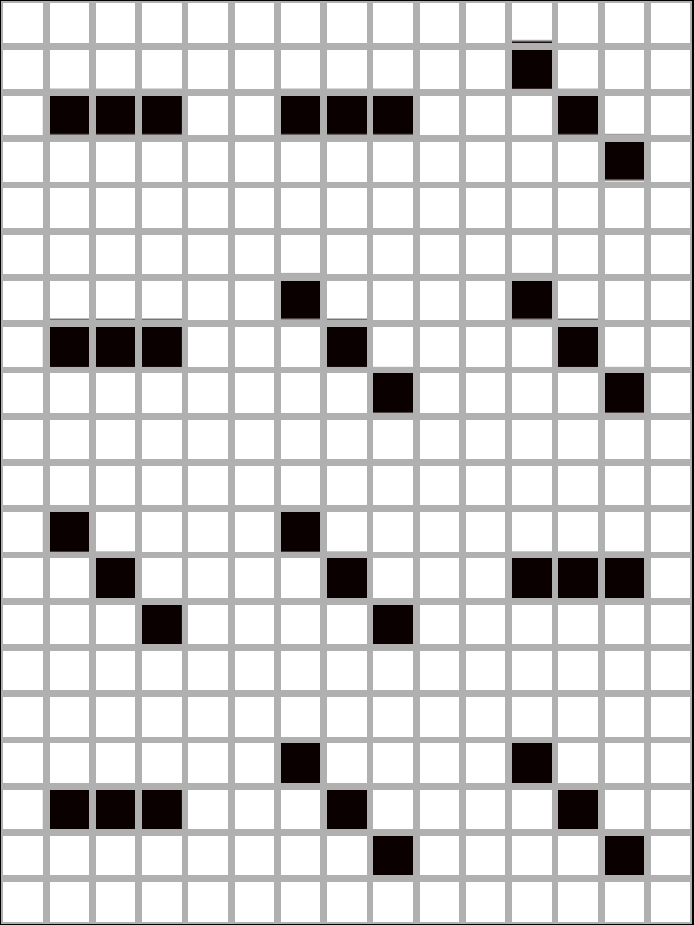}
\\[2ex]
\
\includegraphics[scale=0.4]{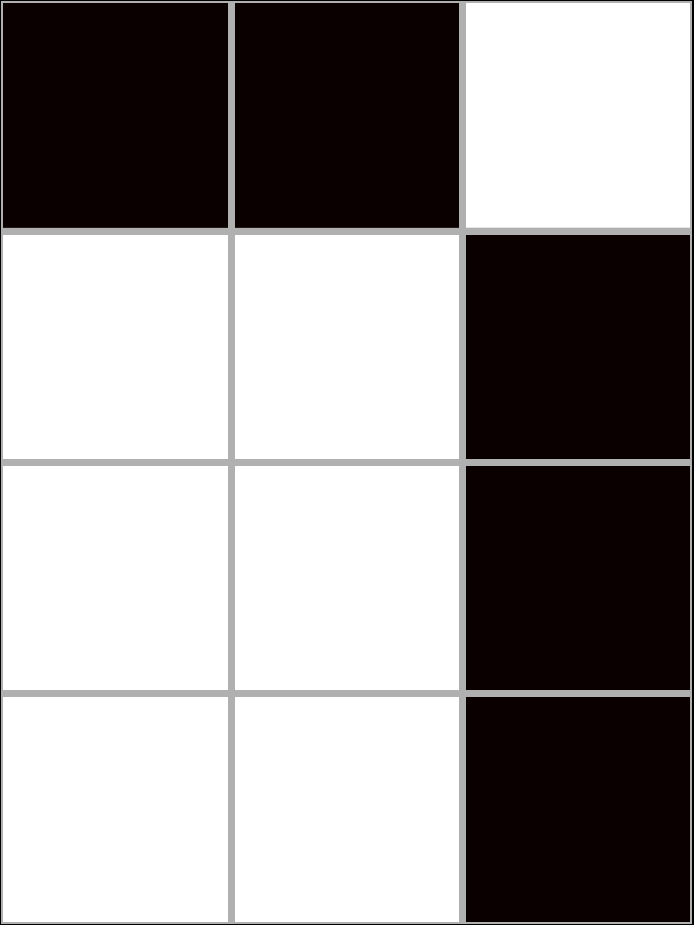} \quad
\includegraphics[scale=0.4]{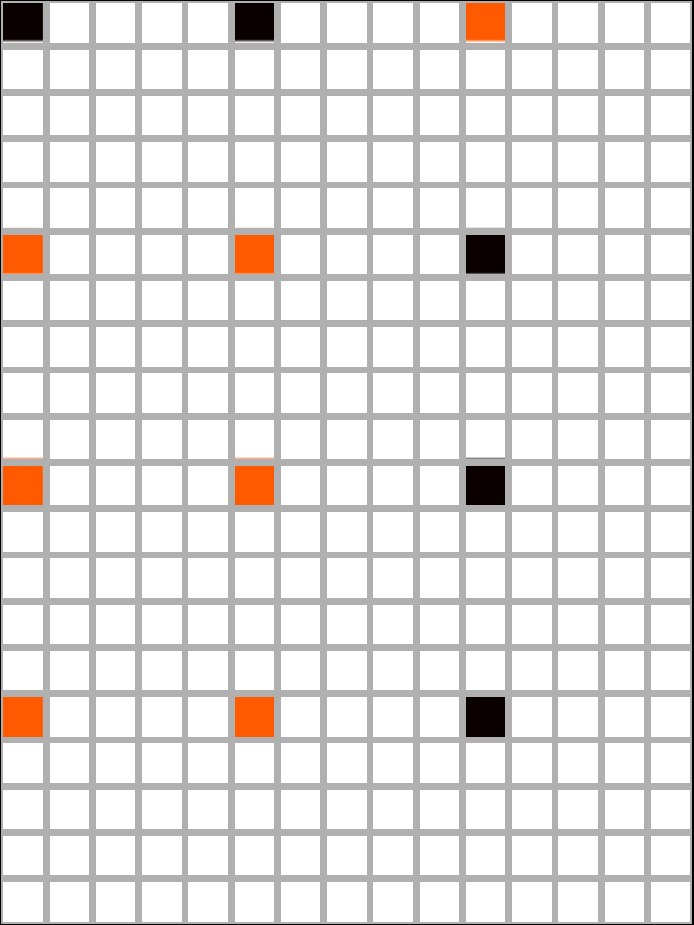} \quad
\includegraphics[scale=0.4]{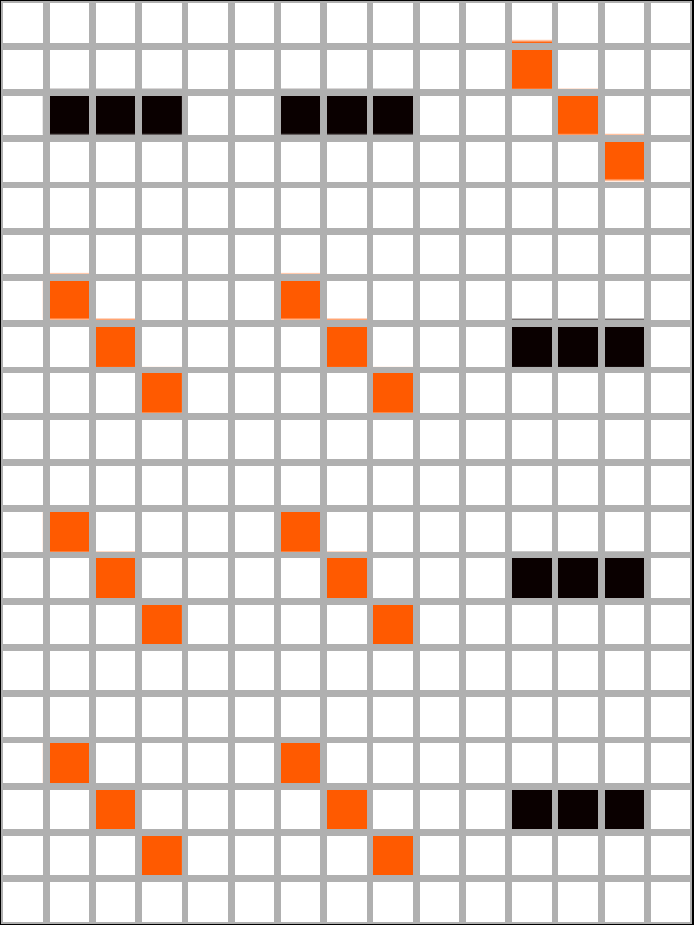} \quad
\includegraphics[scale=0.4]{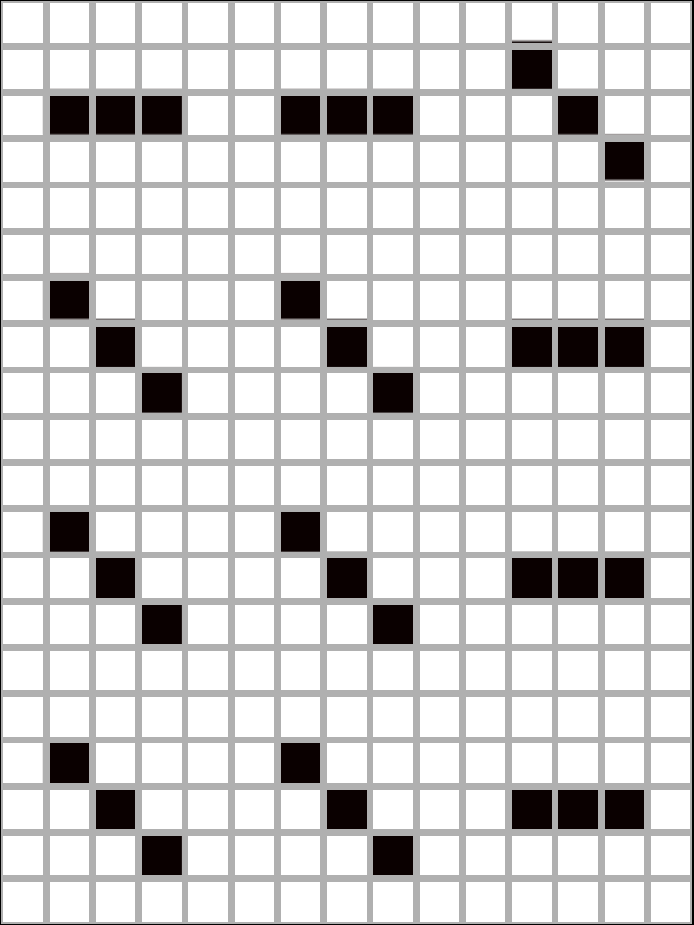}
\caption{Construction of 2PC-equivalent fiber structures based on $\uuS(1) \sim \uuS(2)$ of \autoref{fig_2d_example}}
\label{fig_fiber}
\end{figure}

Finally, it should be noted that the results of the present work may be used in order to generate 2PC-equivalent structures for practical applications, e.g., for fiber- and particle-reinforced materials. Again, starting from the root structures $\uuS(1) \sim \uuS(2)$, inclusion-reinforced 2PC-equivalent structures can be generated. The composition shown in \autoref{fig_fiber} shows how bimodal 2PC-equivalent fiber structure with two different fiber orientations and, strictly speaking, different fiber lengths can be generated. Alternatively, instead of the two-phases structures $\uuS(1) \sim \uuS(2)$ one could begin, e.g., with the generated three-phase structures $\uuS''(1) \sim \uuS''(2)$ displayed in \autoref{fig_2d_example} as parent structures and construct 2PC-equivalent child structures with three different orientations and lengths.

\begin{figure}[H]
\centering
\includegraphics[scale=0.4]{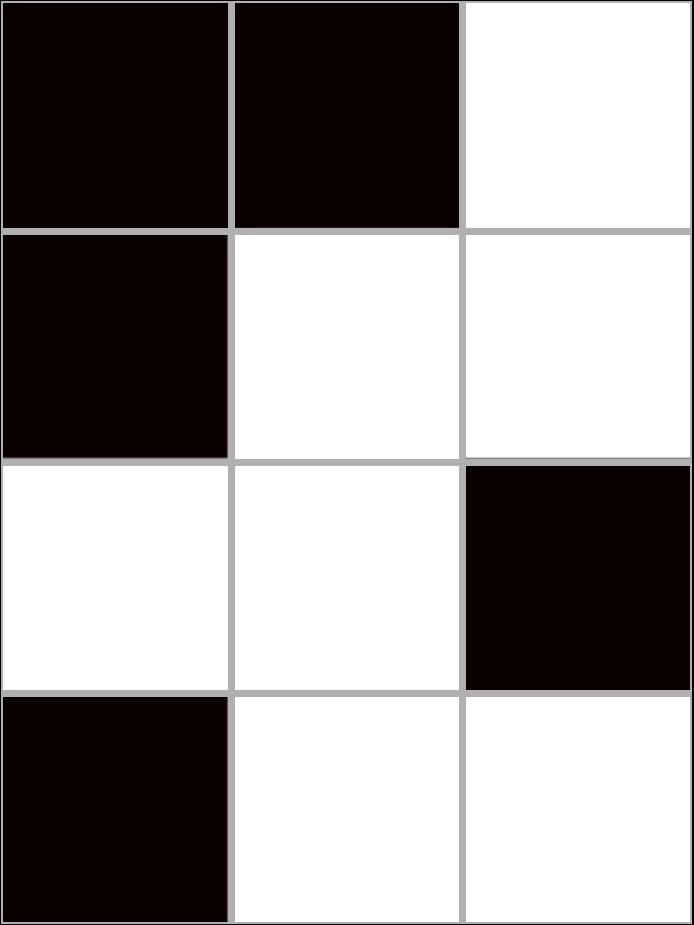} \quad
\includegraphics[scale=0.4]{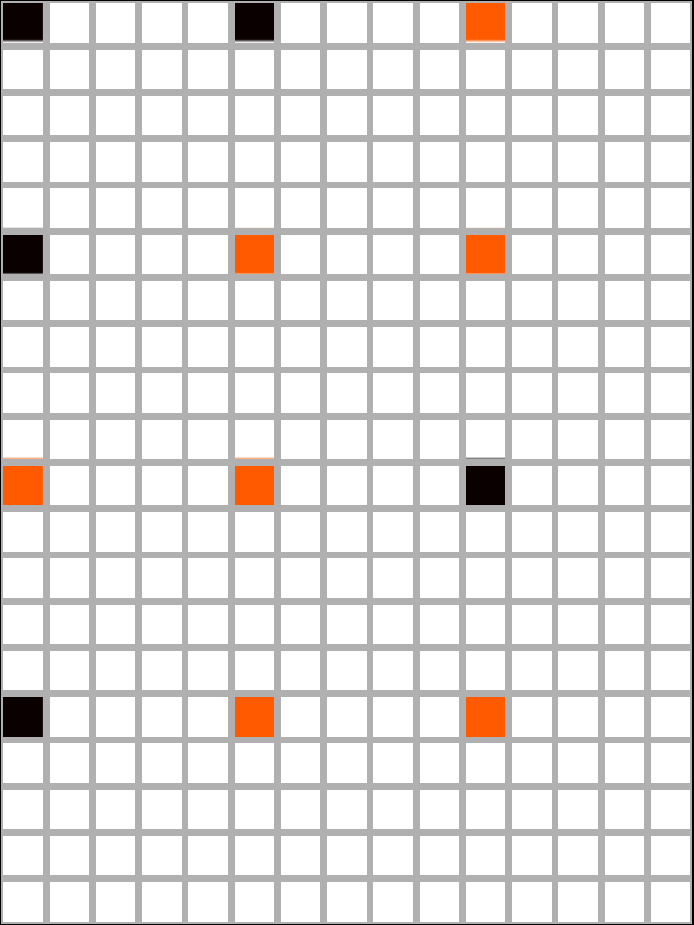} \quad
\includegraphics[scale=0.4]{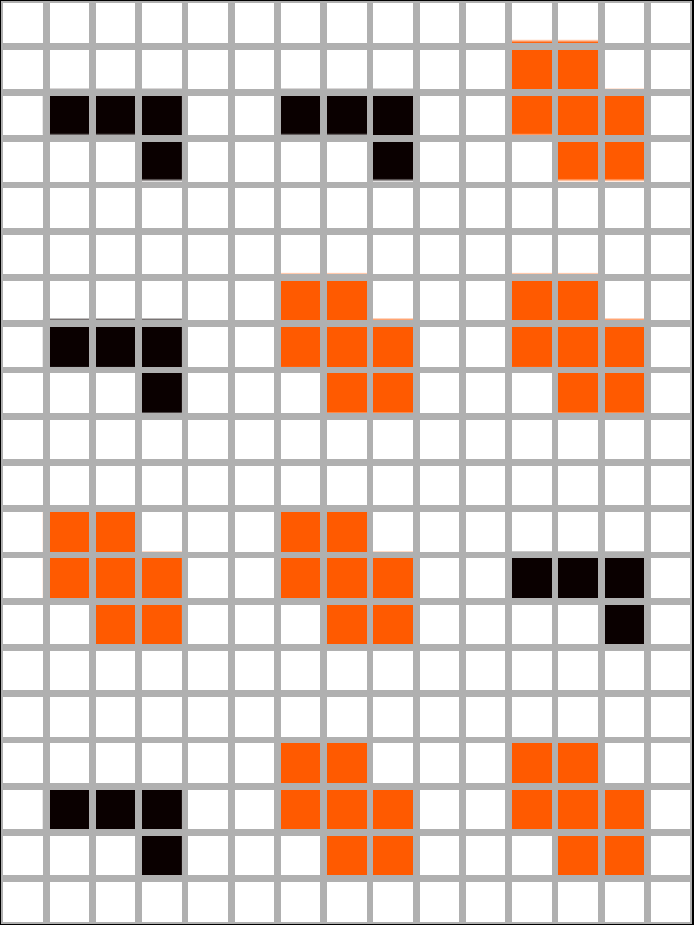} \quad
\includegraphics[scale=0.4]{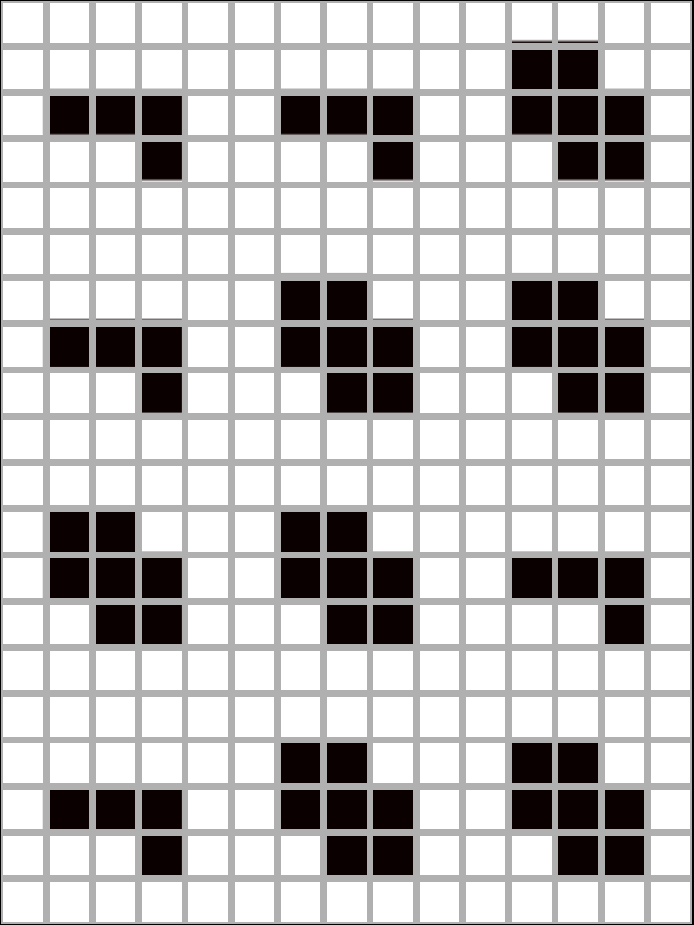}
\\[2ex]
\
\includegraphics[scale=0.4]{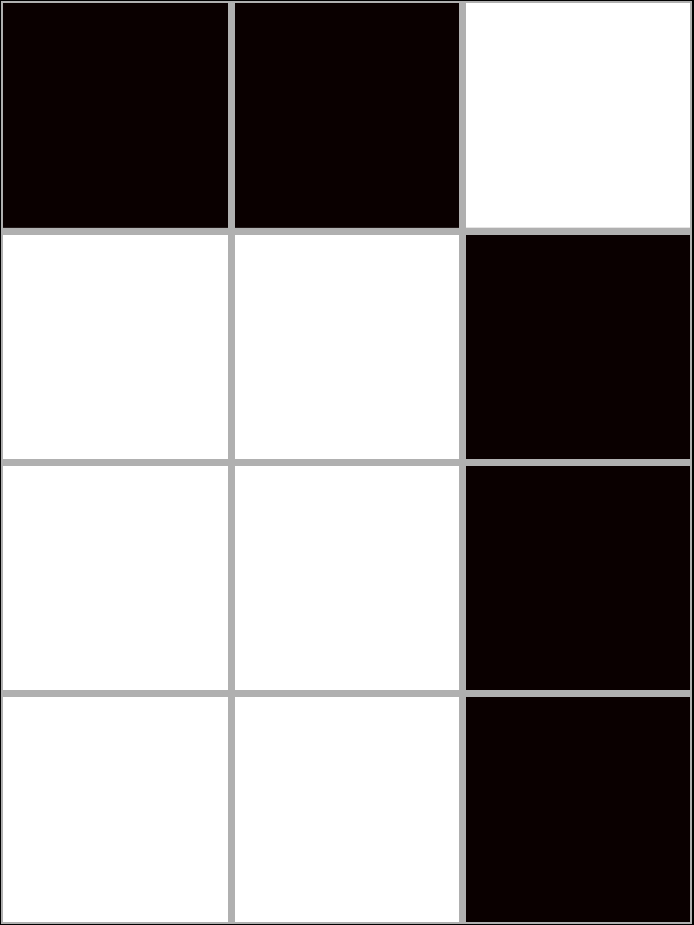} \quad
\includegraphics[scale=0.4]{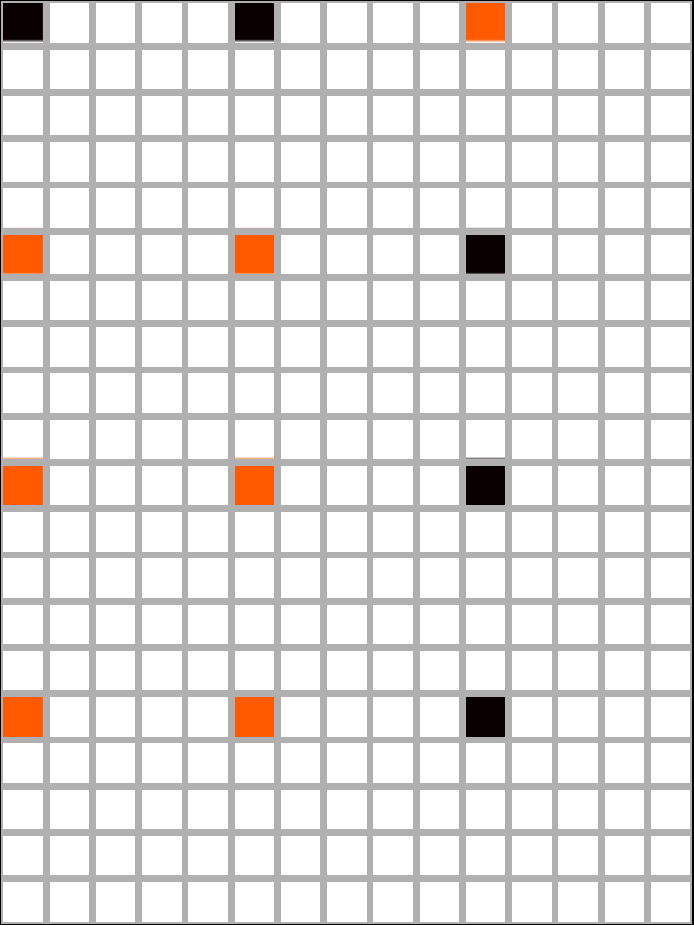} \quad
\includegraphics[scale=0.4]{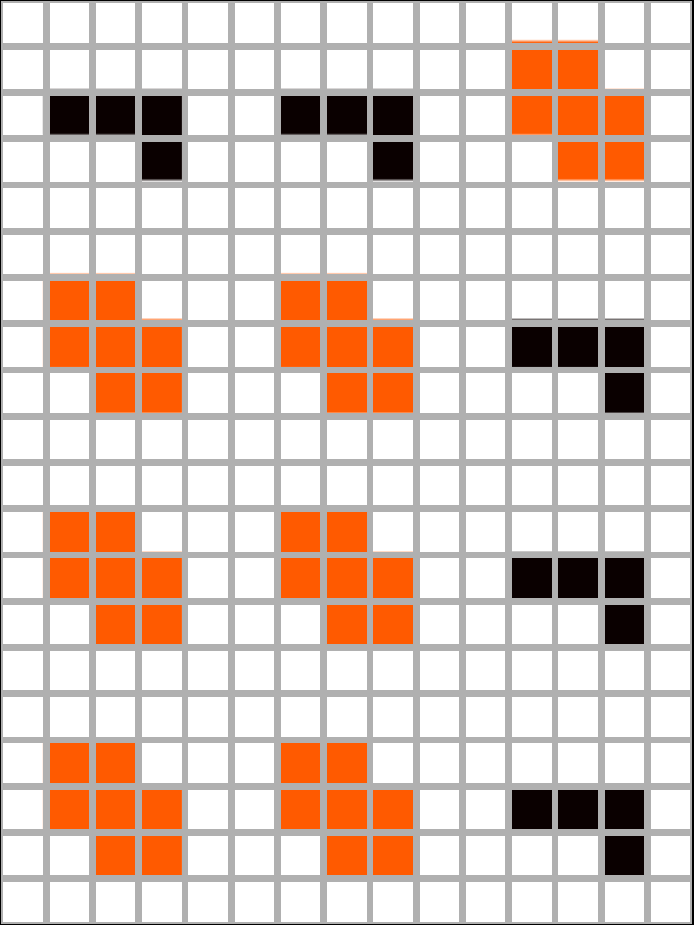} \quad
\includegraphics[scale=0.4]{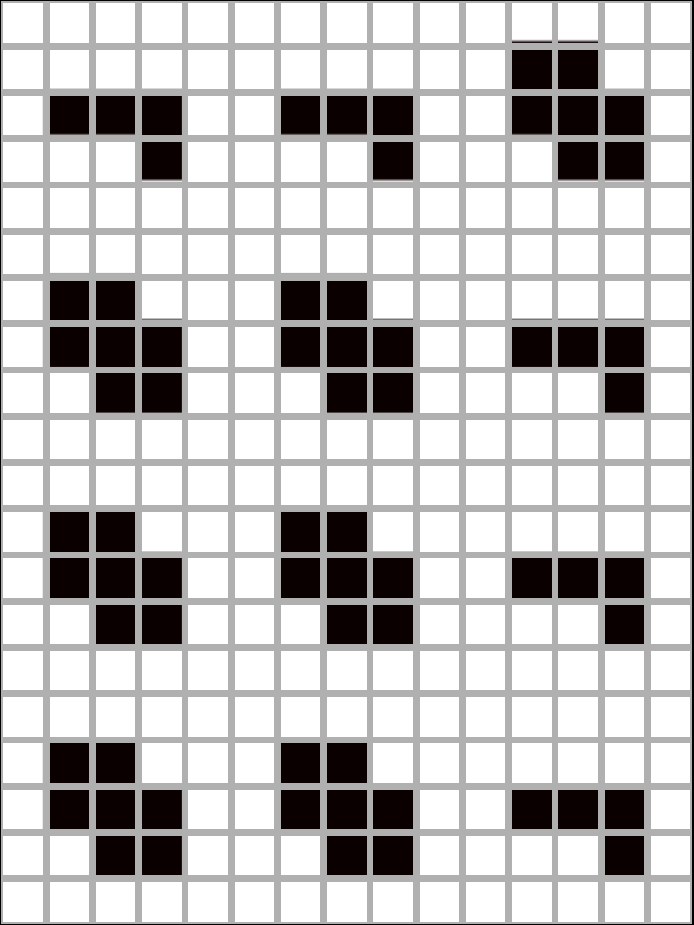}
\caption{Construction of 2PC-equivalent particle structures based on $\uuS(1) \sim \uuS(2)$ of \autoref{fig_2d_example}}
\label{fig_particle}
\end{figure}

Naturally, this strategy can be used for arbitrary particle structures. The composition shown in \autoref{fig_particle} shows the generation of bimodal 2PC-equivalent structures with two different particle shapes. 

\begin{figure}[H]
\centering
\includegraphics[scale=0.4]{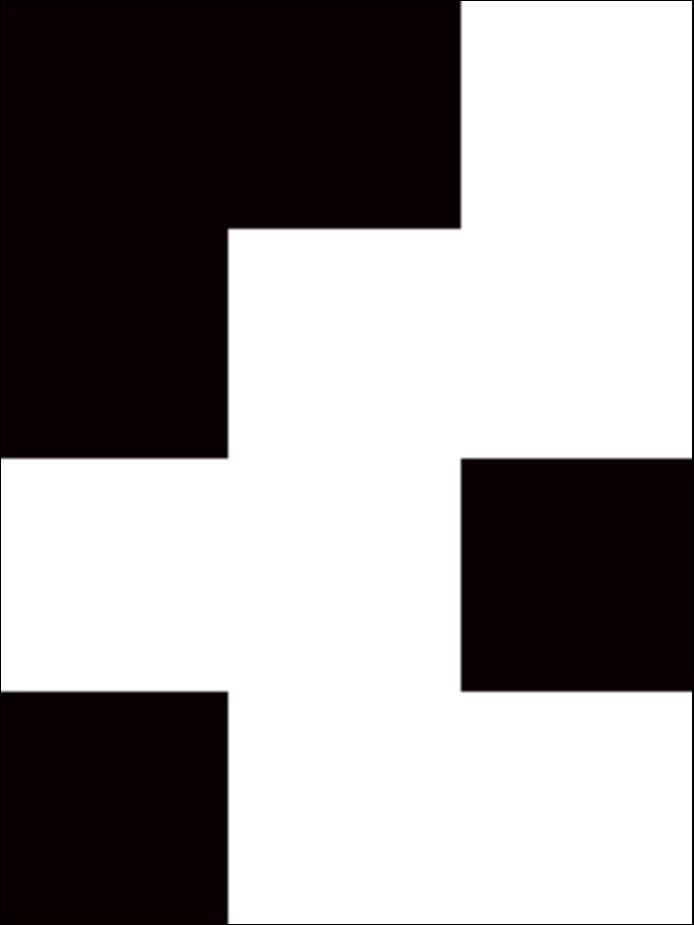} \quad
\includegraphics[scale=0.4]{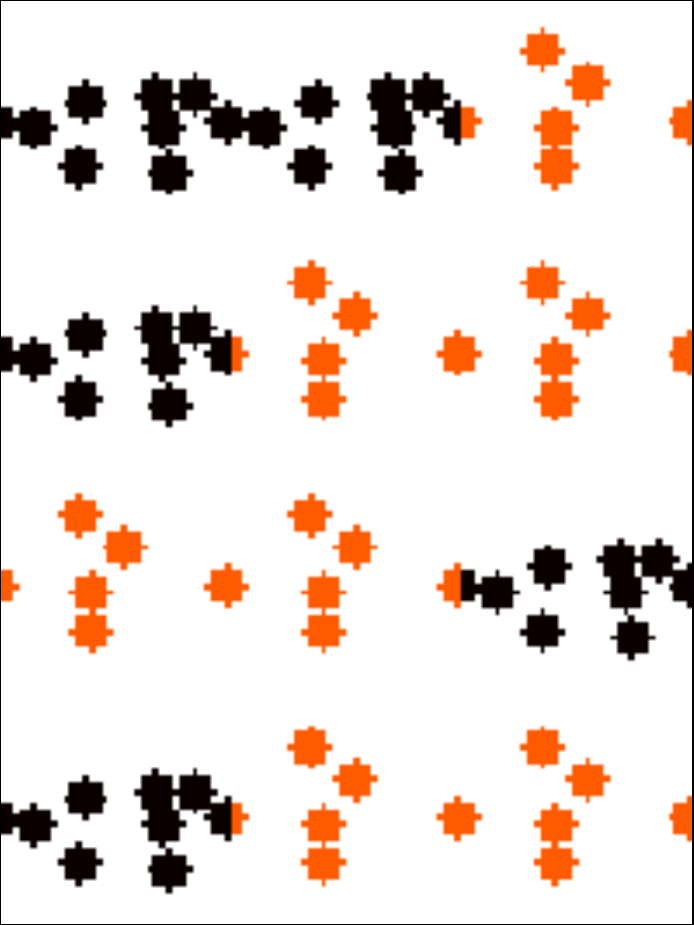}
\\[2ex]
\
\includegraphics[scale=0.4]{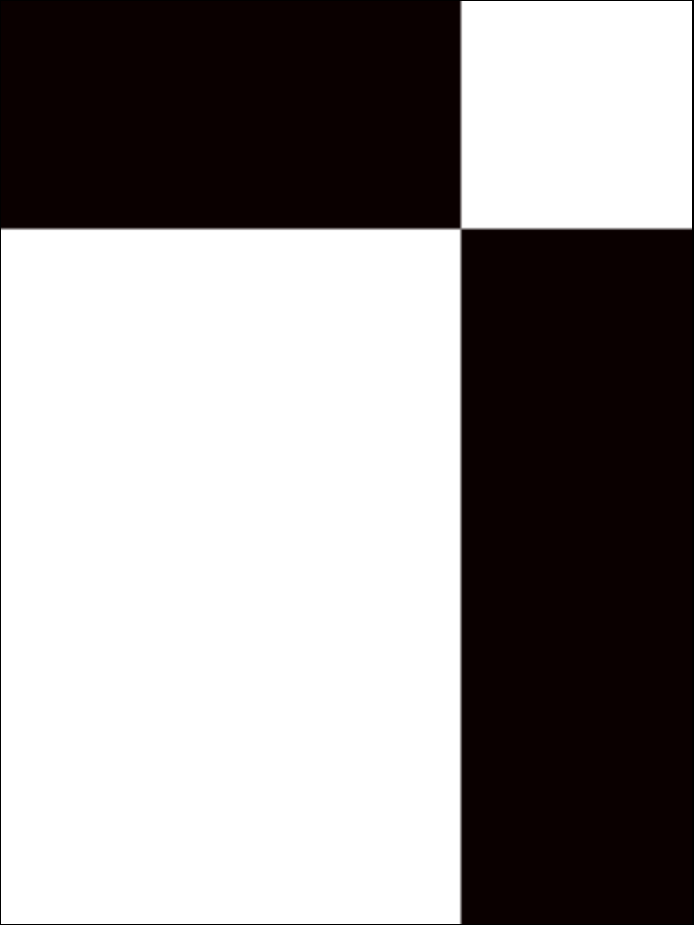} \quad
\includegraphics[scale=0.4]{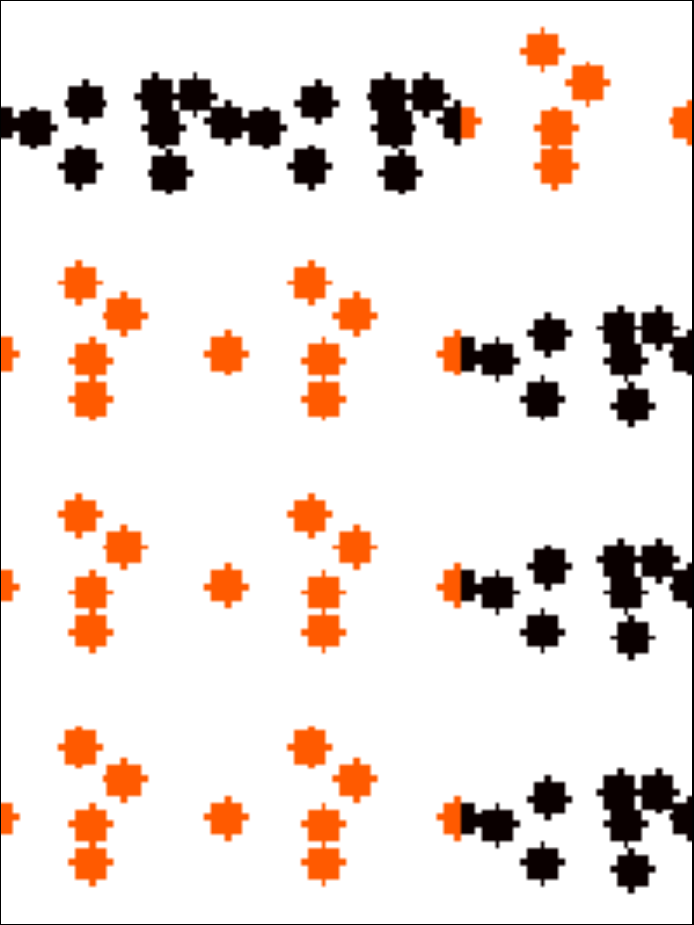}
\caption{Construction of 2PC-equivalent structures based on $\uuS(1) \sim \uuS(2)$ of \autoref{fig_2d_example} and two transition-coherent kernels}
\label{fig_fiber_ca}
\end{figure}

As an option, kernels or binary structures with coherent edges can be used in order form structures with inter-phase coherent transitions. As an example consider \autoref{fig_fiber_ca}, where edge-coherent kernels have been applied. Postponed phase coalescence would then yield new phase distributions with more complex statistics. The structures displayed in \autoref{fig_fiber_ca} can be interpreted in the field of materials science as the cross section of a three-dimensional fiber-reinforced structure with two main fiber bundles. For more complex distributions, one may consider, e.g., the structures $\uuS''(1) \sim \uuS''(2)$ as parent structures in order to generate child structures with three corresponding transition-coherent kernels. In order to generate complex and realistic microstructures with identical 2PC, the use of a set of (compatible) Wang tiles \cite{doskar2014,doskar2018} in combination with several levels of phase extension in order to increase the richness of the overall result is a straight-forward offering.

Finally, it should be remarked that applications needing a high-number of phases can generate 2PC-equivalent structures by applying the phase extension operation repeatedly. As an example from materials science, polycrystals can be considered as structures with a high number of phases, where each phase corresponds to a single grain. The root structures $\uuS(1) \sim \uuS(2)$, for example, can be used to generated 2PC-equivalent polycrystals with arbitrary number of grains/phases. 

\begin{figure}[H]
\centering
\includegraphics[width=0.9\textwidth]{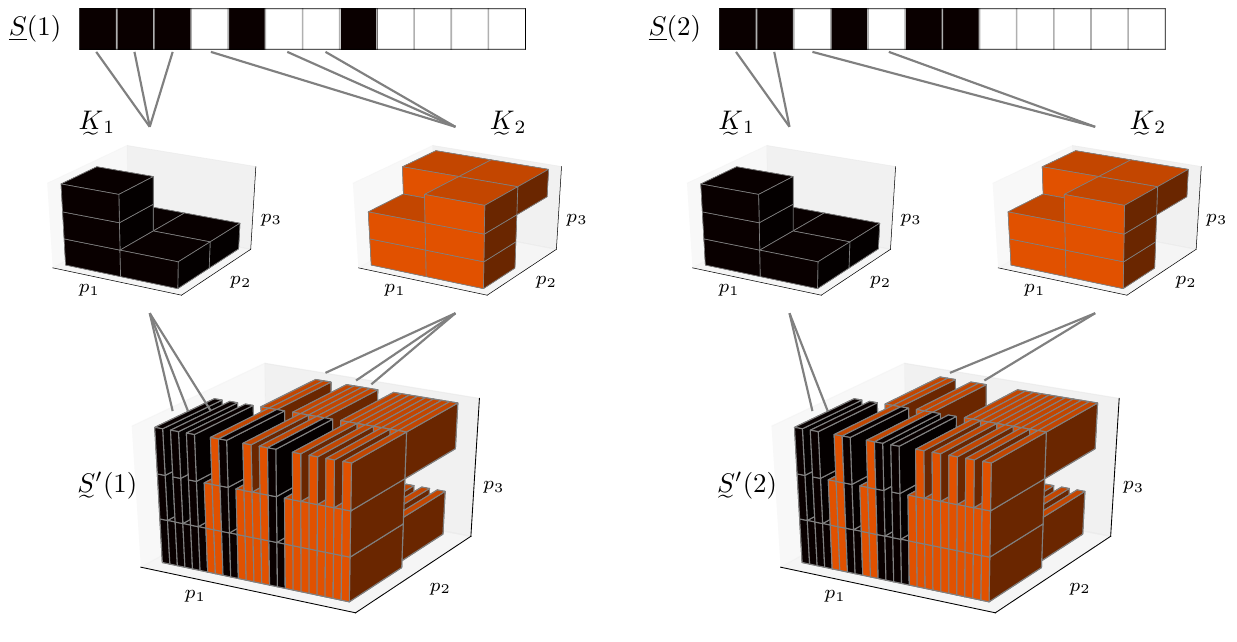}
\caption{Kernel-based extension: 
root one-dimensional structures $S(1) \sim S(2)$, kernels $\protect\utilde{K}_1$ and $\protect\utilde{K}_2$ for the extension of phases 1 and 2, and extended structures $\protect\utilde{S}'(1) \sim \protect\utilde{S}'(2)$.
}
\label{fig_ex_3d_kbe}
\end{figure}

\paragraph{Three-dimensional example.}
Consider again the root one-dimensional structures $\uS(1) \sim \uS(2)$ given by \eqref{eq_1d_Isum12} and illustrated in \autoref{fig_ex_1d}. We now extend these structures based on the kernels $\utK_1, \utK_2 \in \bbI^{\uz}$ for $\uz = (2,2,3)$ 
\begin{eqnarray}
	K_{1,p_1p_2p_3} 
	&=&
	\begin{cases}
	1 & (p_1,p_2,p_3) \in \{
	(0,0,0),(0,0,1),(0,0,2)
	,(0,1,0)
	,(1,0,0)
	,(1,1,0)
	\}
	\\
	0 & \text{else}
	\end{cases}
	\ ,
	\\
	K_{2,p_1p_2p_3} 
	&=&
	\begin{cases}
	1 & (p_1,p_2,p_3) \in \{
	(0,0,0),(0,0,1)
	,(0,1,0),(0,1,2)
	,(1,0,0),(1,0,1),(1,0,2)
	,(1,1,2)
	\}
	\\
	0 & \text{else}
	\end{cases}
	\ .
\end{eqnarray}
The root one-dimensional structures are again displayed in the top of \autoref{fig_ex_3d_kbe}, while the corresponding kernel-based extended structures are illustrated at the bottom of \autoref{fig_ex_3d_kbe}.

It should be stressed that this example illustrates not only a three-dimensional example but also the possibility to dimension-extend any given set of 2PC-equivalent structures to arbitrary dimensions, sizes and aspect ratios. 

\subsection{Effective linear properties in homogenization theory of periodic media}

In this section a structure is interpreted as a microscopic unit cell of a periodic medium. We consider the two-dimensional periodic conductivity problem in a structure domain $S \subset \bbR^2$ with phase-wise constant conductivity $\uuK(\ux) \in \bbR^{2 \times 2}$
\begin{equation}
	\mathrm{div}(\uuK(\ux) \ \mathrm{grad}(u(\ux))) = 0
	\quad \ux \in S
	\ , \quad
	u(\ux) = \bar{\ug}^T \ux + v(\ux) 
	\quad \ux \in \partial S
\end{equation}
where, for given $\bar\ug$, the solution $u(\ux)$ is a superposition of the linear field $\bar\ug^T\ux$ and a fluctuation $v(\ux)$, which is periodic over the structure. The flux in the structure
\begin{equation}
	\uf(\ux) = \uuK(\ux) \mathrm{grad}(u(\ux))
	\label{eq_fKg_micro}
\end{equation}
is of interest. In homogenization theory of periodic media it is well-known that the effective gradient of the solution evaluates due to the periodicity of $v(\ux)$ to the given $\bar\ug$ 
\begin{equation}
	\frac{1}{|S|} \int_S \mathrm{grad}(u(\ux)) \ \rmd S = \bar\ug
\end{equation}
and that the volume average of the flux, i.e., the effective flux, $\bar\uf$ is linear in the effective gradient $\bar\ug$. Therefore, the representation
\begin{equation}
	\bar\uf = \bar\uuK \ \bar\ug
\end{equation}
exists. The constant $\bar\uuK$ is referred to as the effective conductivity of the structure and characterizes the macroscopic constitutive behavior of the periodic material inherited from the microscopic one \eqref{eq_fKg_micro}. If the solution $u(\ux)$ is considered as the temperature field, then $\bar\uuK$ corresponds to the negative of the effective heat conductivity, while for $u(\ux)$ being the electric potential $\bar\uuK$ reflects the effective electric conductivity. 

Many homogenization schemes target an approximation or bounds of the effective material properties, here in terms of $\bar\uuK$. These are often built upon the material properties of the constituents and statistical properties of the structure. The vast majority of such schemes rely on one-point statistical information, i.e., volume fractions of the constituent materials. The 2PC of the present work corresponds to the two-point statistics of the microstructure. Homogenization schemes based on the 2PC are expected to be vastly superior to ones based on volume fractions. But what if the 2PC is not enough information? The investigations of the present works motivate the question: can two or more 2PC-equivalent microstructures have large deviations in the respective $\bar\uuK$? This would then imply that corresponding homogenization schemes based on the 2PC have from the very beginning no chance to accurately approximate $\bar\uuK$ for all of these 2PC-equivalent structures. 

As a first example, we consider again the two-dimensional structures $\uuS(1) \sim \uuS(2)$ presented in \autoref{fig_2d_example}. We consider the phase-wise constant conductivity field for isotropic constituents
\begin{equation}
	\uuK(\ux) 
	= 
	k(\ux) \uuI
	\ , \quad
	k(\ux)
	=
	\begin{cases}
	k_1 & \ux \in \text{ phase 1 } \\
	k_2 & \ux \in \text{ phase 2 }
	\end{cases}
	\label{eq_K}
\end{equation}
For the current case of isotropic two-phase materials, bounds on the effective conductivity are well-known in homogenization theory. Based purely on the volume fraction $v_1 \in [0,1]$ of phase 1, the upper and lower so-called first order bounds can be computed
\begin{equation}
	\bar\uuK^{1+} = [v_1 k_1 + (1-v_1) k_2]\uuI
	\ , \quad
	\bar\uuK^{1-} = \left[v_1 \frac{1}{k_1} + (1-v_1) \frac{1}{k_2}\right]^{-1}\uuI
	\ ,
	\label{eq_vr}
\end{equation}
where $\bar\uuK^{1+}$ is referred to in the literature as the Voigt bound, while $\bar\uuK^{1-}$ is referred to as the Reuss bound, see, e.g., \cite{Voigt1910}, \cite{Reuss1929} or \cite{Willis1981}. The Voigt and Reuss bounds given in \eqref{eq_vr} bound the effective conductivity from below and above via
\begin{equation}
	\bar\ug^T \bar\uuK^{1-} \bar\ug 
	\leq
	\bar\ug^T \bar\uuK \ \bar\ug 
	\leq 
	\bar\ug^T \bar\uuK^{1+} \bar\ug 
	\quad
	\forall \bar\ug .
	\label{eq_bounds_ineq}
\end{equation}
This then implies, e.g., 
\begin{equation}
	\bar{K}^{1-}_{ii} 
	\leq 
	\bar{K}_{ii} 
	\leq 
	\bar{K}^{1+}_{ii} 
	\ , \quad
	i = 1,2.
\end{equation}
More elaborated bounds can be constructed based on higher-order correlation information of the microstructure. The so-called Hashin-Shtrikman bounds, see, e.g., \cite{Willis1981} or \cite{Torquato2002}, are second-order bounds based on the two-point correlation. These bounds are usually simplified by assuming no long-range order and isotropic two-point correlation. For the two-dimensional conductivity problem at hand these simplifications yield
\begin{align}
	\bar{\uuK}^{2+}
	&= \left[\left( v_1 \frac{1}{k_\mathrm{max} + k_1} + (1-v_1) \frac{1}{k_\mathrm{max} + k_2} \right)^{-1} - k_\mathrm{max} \right] \uuI
	\ , \nonumber\\
	\bar{\uuK}^{2-}
	&= \left[\left( v_1 \frac{1}{k_\mathrm{min} + k_1} + (1-v_1) \frac{1}{k_\mathrm{min} + k_2} \right)^{-1} - k_\mathrm{min} \right] \uuI
	\ , \nonumber\\
	k_\mathrm{min}
	&= \min\{k_1,k_2\}
	\ , \nonumber\\
	k_\mathrm{max}
	&= \max\{k_1,k_2\}
	\ ,
	\label{eq_hs}
\end{align}
see \cite{Torquato2002}. Strictly speaking, since the structures of the present work do \emph{not} show isotropic 2PC, the Hashin-Shtrikman bounds are not valid, i.e., analogous relations to \eqref{eq_bounds_ineq} do not hold already if the 2PC is not isotropic. For some structures, the corresponding 2PC may be considered as approximately isotropic and to show approximately no long-range order, such that the Hashin-Shtrikman bounds \eqref{eq_hs} can be considered as bounds in the corresponding approximate sense. 

\begin{figure}[H]
\centering
\includegraphics[scale=1.2]{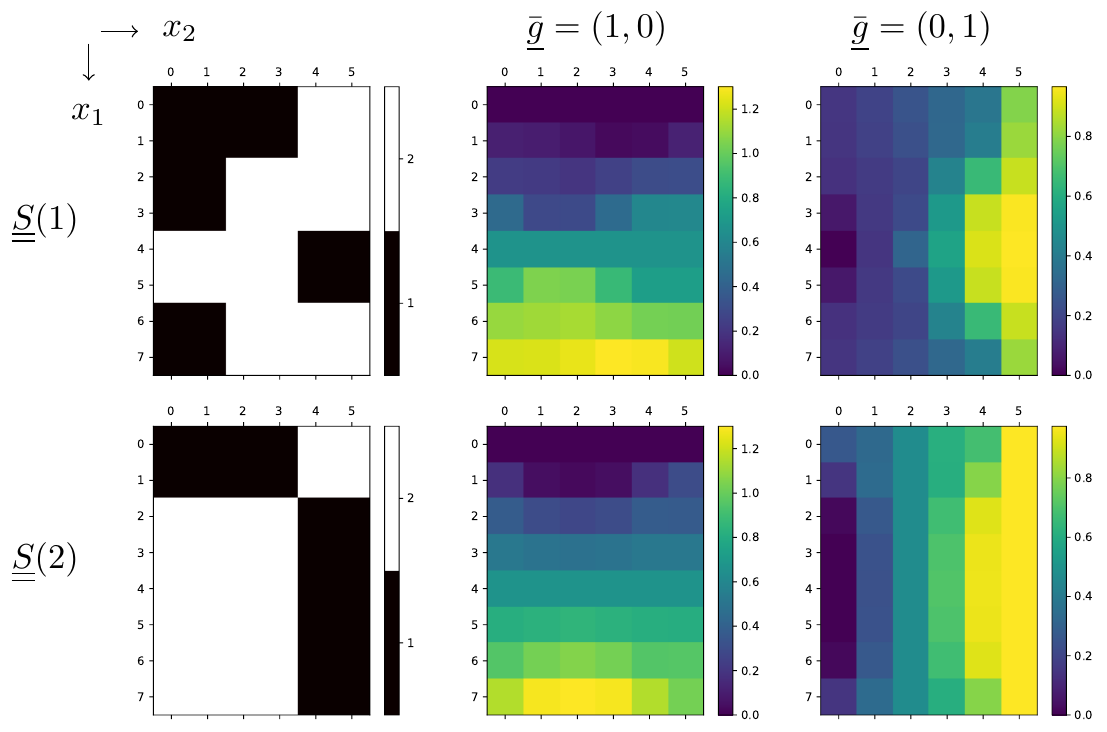}
\\
(a) 
\\[2ex]
\includegraphics[scale=1.2]{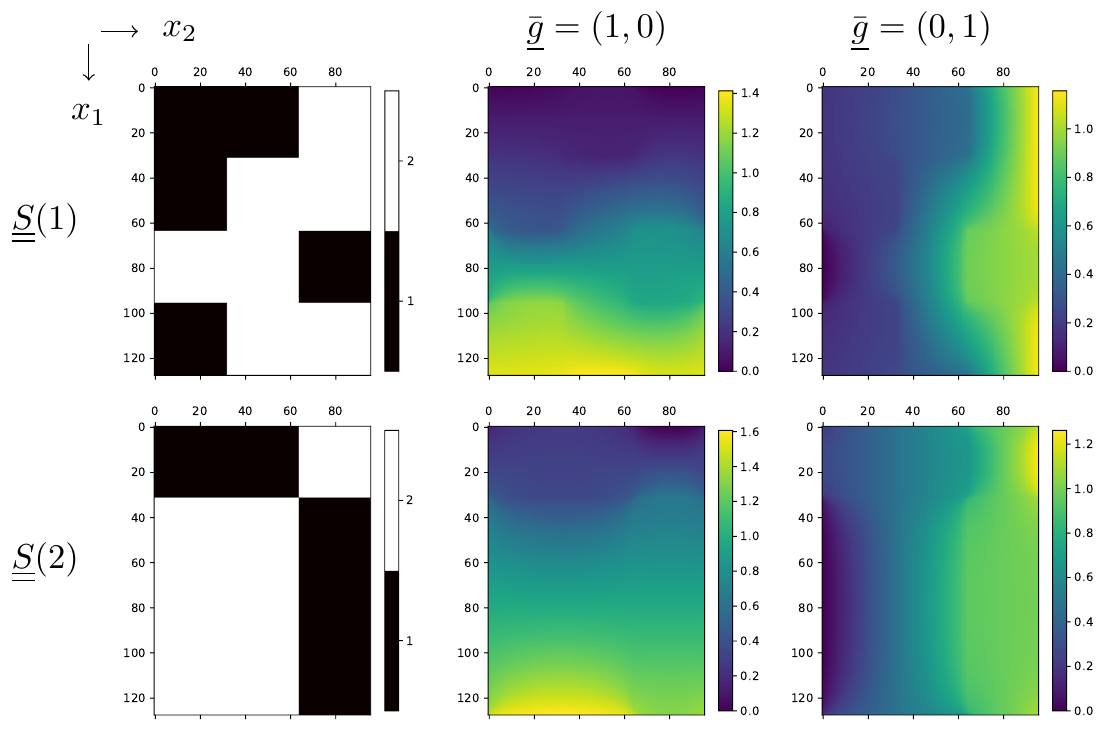}
\\
(b) 
\caption{Structures $\uuS(1)$ and $\uuS(2)$ displayed in \autoref{fig_2d_example}: (a) with 2x resolution (left plots) and corresponding solution fields $u(\ux)$ for $\bar\ug = (1,0)$ (middle plots) and $\bar\ug = (0,1)$ right plots; (b) with 32x resolution (left plots) and corresponding solution fields $u(\ux)$ for $\bar\ug = (1,0)$ (middle plots) and $\bar\ug = (0,1)$ right plots.}
\label{fig_h_2}
\end{figure}

The two-dimensional homogenization problem for conductivity has been solved numerically for $k_1 = 10$ and $k_2 = 1$ with the FANS approach of \cite{Leuschner2018a} for 2x resolution and 32x resolution of the structures $\uuS(1) \sim \uuS(2)$, see \autoref{fig_h_2} (a) and (b). The oversampling ratios of 2 and 32 have been considered in order to rule out discretization issues.

\begin{figure}[H]
\centering
\includegraphics[width=0.45\textwidth]{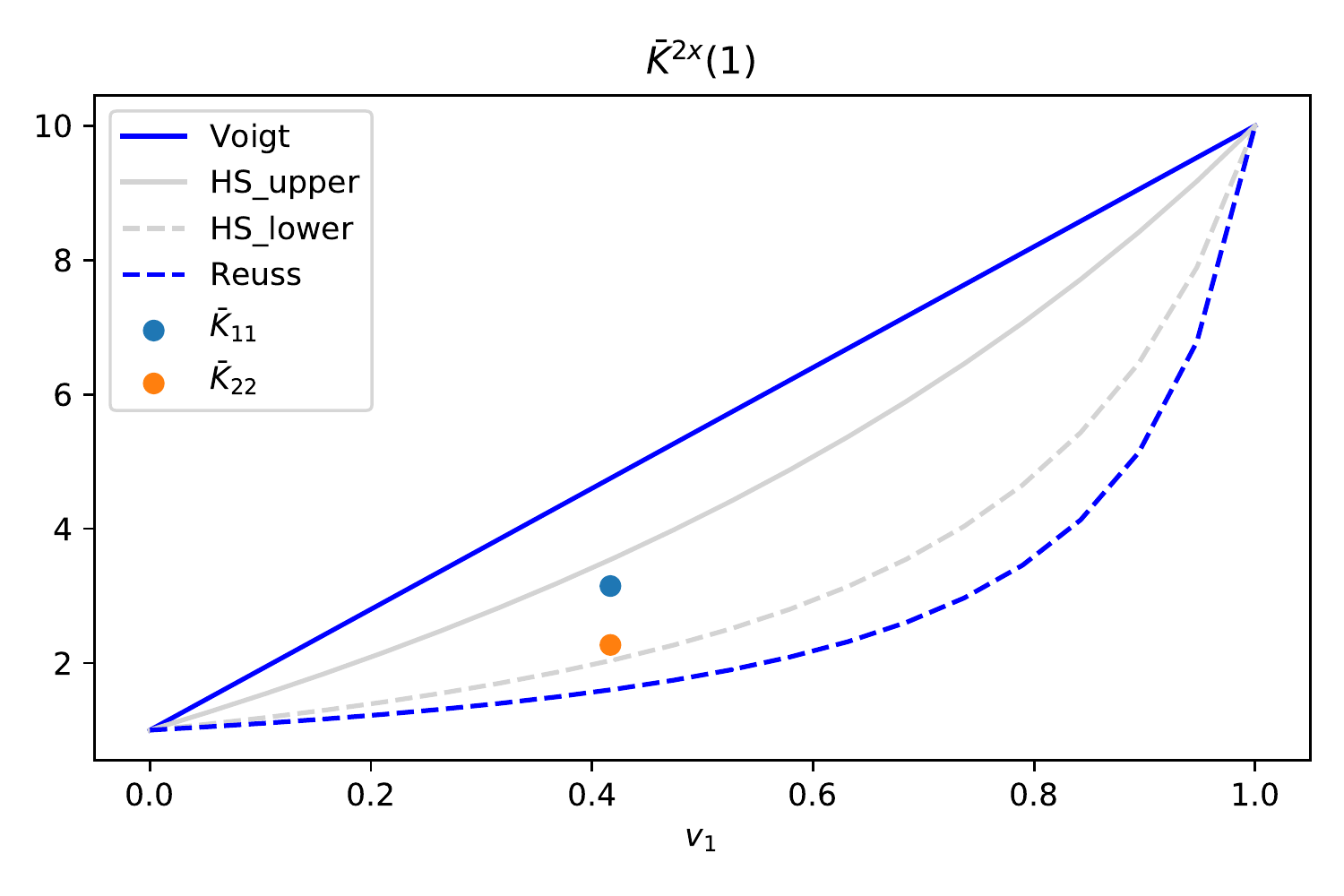}
\includegraphics[width=0.45\textwidth]{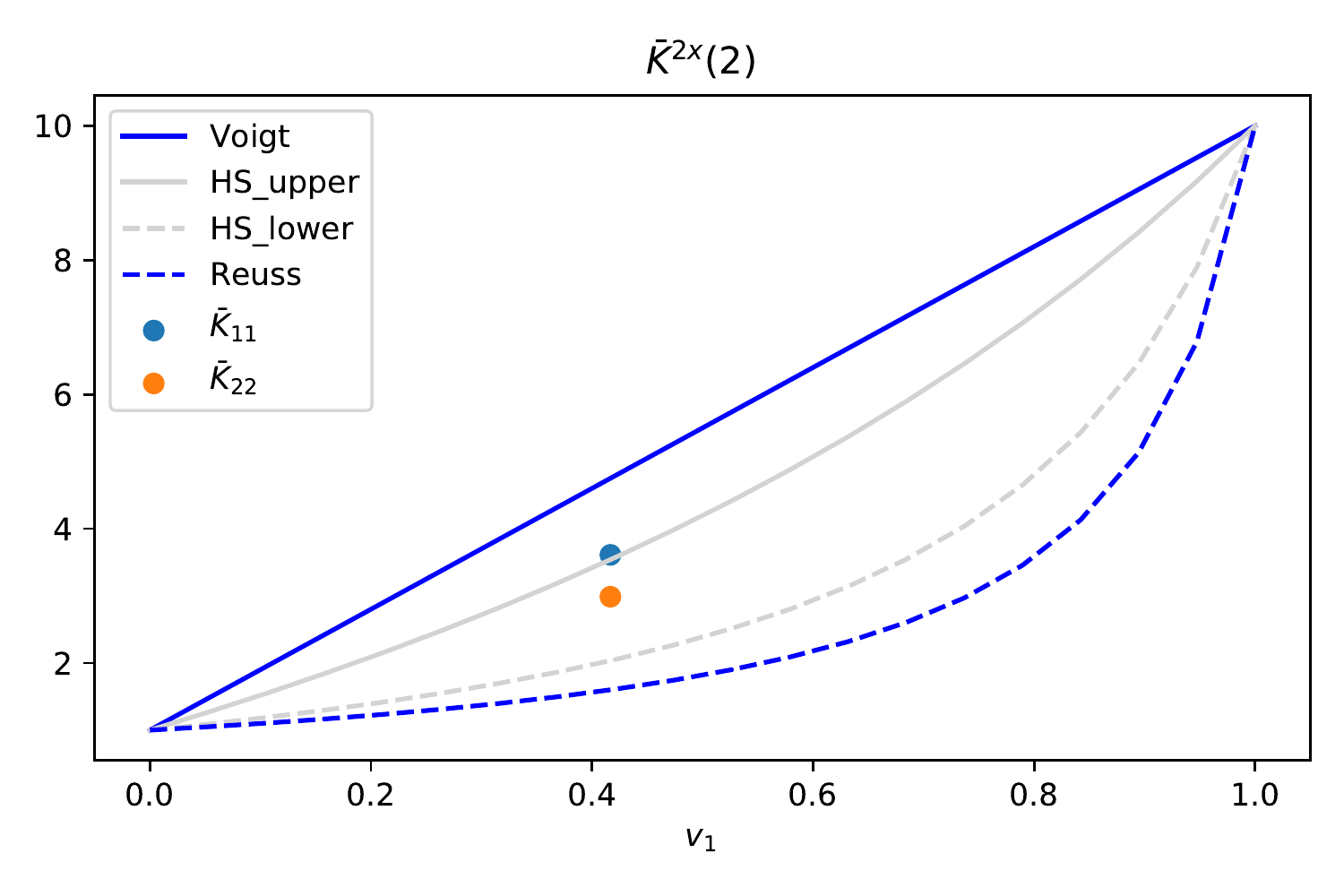} \\
\includegraphics[width=0.45\textwidth]{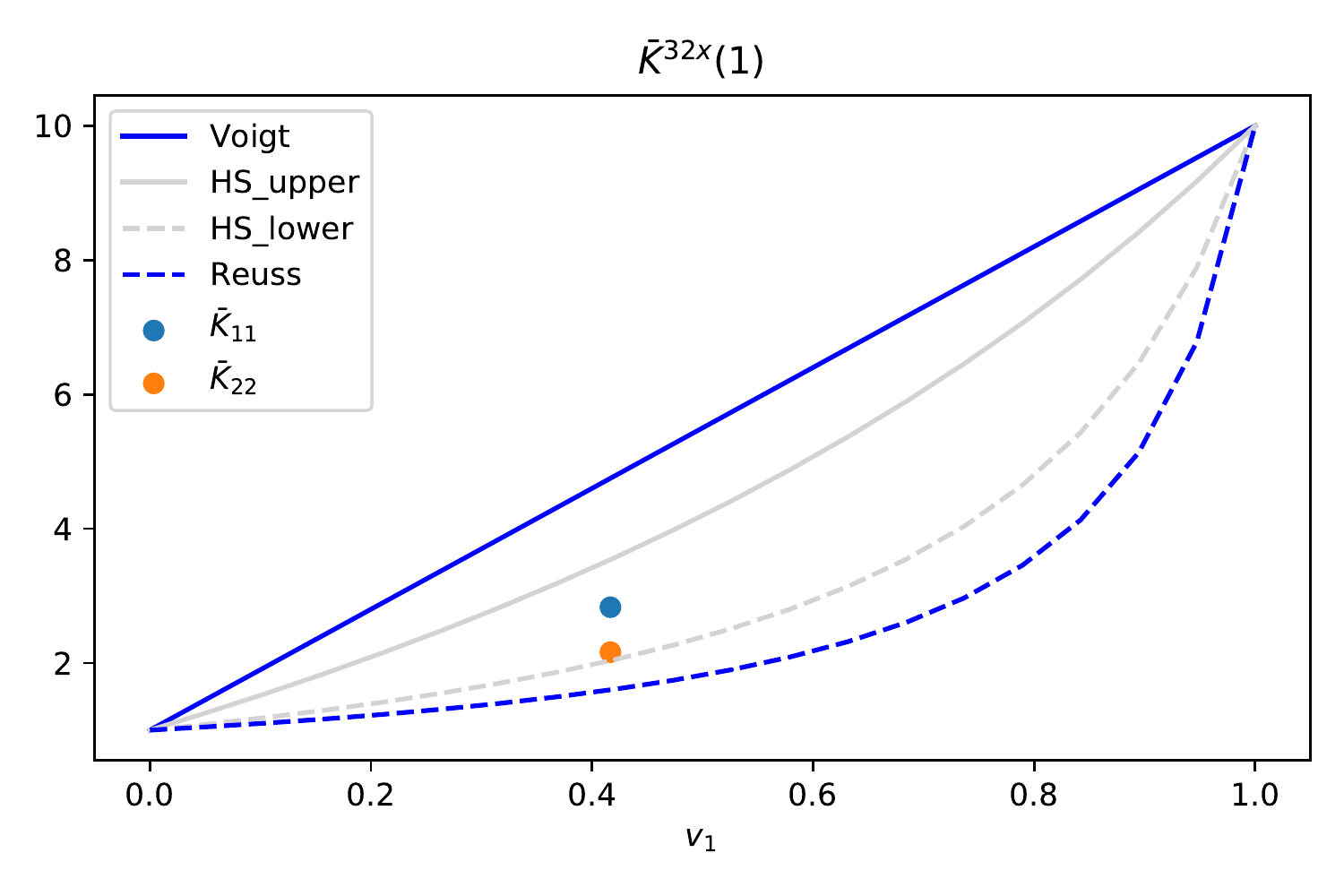}
\includegraphics[width=0.45\textwidth]{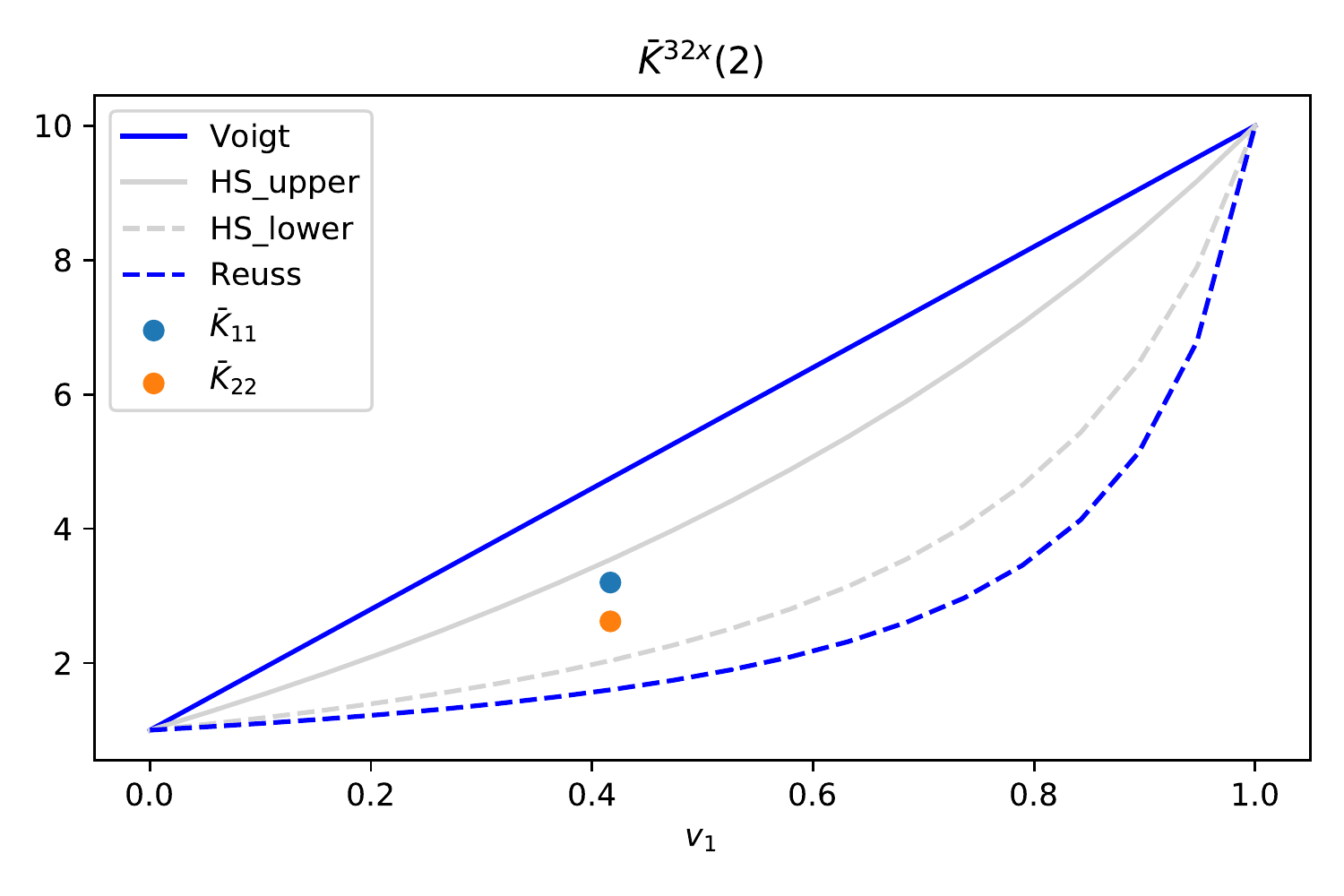}
\caption{Evaluation of Voigt, Reuss and Hashin-Shtrikman bounds for $v_1 \in [0,1]$ with $k_1 = 10$ and $k_2 = 1$; the effective conductivites of the structures $\uuS'''(1) \sim \uuS'''(2)$ for 2x and 32x resolution is depicted by the points at $v_1 \approx 0.42$: the $\bar{K}_{11}$ component of the corresponding effective conductivity is displayed by the corresponding blue point, while the $\bar{K}_{22}$ component is depicted by the corresponding orange point.}
\label{fig_bounds_1}
\end{figure}

For the 2x resolution of the structures, the corresponding effective conductivities evaluate to
\begin{equation}
	\bar\uuK^{\text{2x}}(1)
	=
	\begin{pmatrix}
	3.1455 & 0 \\
	0 & 2.2687
	\end{pmatrix}
	\ , \quad
	\bar\uuK^{\text{2x}}(2)
	=
	\begin{pmatrix}
	3.6099 & 0 \\
	0 & 2.9865
	\end{pmatrix}
	\ .
\end{equation}
For the 32x resolution of the structures, the corresponding effective conductivities evaluate to
\begin{equation}
	\bar\uuK^{\text{32x}}(1)
	=
	\begin{pmatrix}
	2.8305 & 0 \\
	0 & 2.1615
	\end{pmatrix}
	\ , \quad
	\bar\uuK^{\text{32x}}(2)
	=
	\begin{pmatrix}
	3.1981 & 0 \\
	0 & 2.6198
	\end{pmatrix}
	\ .
\end{equation}
The volume fraction of phase 1 equals $v_1 = 5/12 \approx 42 \%$. Based on $k_2 < k_1$ and $v_1 < 1/2$, the structures $\uuS(1) \sim \uuS(2)$ can be interpreted as reinforced structures of a material with conductivity $k_2$. Evaluation of the Voigt, Reuss and Hashin-Shtrikman bounds is illustrated in \autoref{fig_bounds_1}. As remarked in the introduction of the bounds, only the Voigt and Reuss bounds are definite bounds for the considered structures, the Hashin-Shtrikman bounds are not valid, strictly speaking, since the structures do not posses the corresponding statistical properties. 

The relative deviations of the effective conductivity obtained by computational homogenization are
\begin{equation}
	\frac{\norm{\bar\uuK^{\text{2x}}(1) - \bar\uuK^{\text{2x}}(2)}}{\norm{\bar\uuK^{\text{2x}}(1)}}
	= 22.05 \%
	\ , \quad
	\frac{\norm{\bar\uuK^{\text{32x}}(1) - \bar\uuK^{\text{32x}}(2)}}{\norm{\bar\uuK^{\text{32x}}(1)}}
	= 
	16.50 \%
	\ .
\end{equation}
This example shows that even though some structures may have identical 2PC the effective properties can still strongly deviate from structure to structure. This naturally implies, in principle, that the accuracy of many homogenization schemes based not only on one- but also on two-point-statistics may not be sufficient for the application at hand. The results of the present work offer benchmark structures for such homogenization schemes. 

It should be remarked that the 2PC equivalent structures $\uuS(1)$ and $\uuS(2)$ differ with respect to the 3PC, cf. \eqref{eq_mpc}. Computation of the 3PC of phase 1 for the corresponding structures and resolutions yields
\begin{equation}
	\frac{ 
	\norm{\utC^\text{2x}_{\langle 3 \rangle 111}(1) - \utC^\text{2x}_{\langle 3 \rangle 111}(2)}
	}
	{
	\norm{\utC^\text{2x}_{\langle 3 \rangle 111}(1)}	
	}
	= 31.88\%
	\ , \quad
	\frac{ 
	\norm{\utC^\text{32x}_{\langle 3 \rangle 111}(1) - \utC^\text{32x}_{\langle 3 \rangle 111}(2)}
	}
	{
	\norm{\utC^\text{32x}_{\langle 3 \rangle 111}(1)}	
	}
	= 24.36 \% 
	\ .
\end{equation}
The deviation in the 3PC for the structures $\uuS(1) \sim \uuS(2)$ indicates that even for relative fine discretization --- as the one for 32x resolution --- a clear statistical difference can be found between $\uuS(1)$ and $\uuS(2)$. Homogenization schemes aiming for the effective properties of these structures may have to include the 3PC or even higher correlations for linear and, probably, also for nonlinear material behavior in order to make reliable predictions in the utmost general scenarios. 

The single pixel flip difference between the 2PC-equivalent structures $\uuS(1)$ and $\uuS(2)$ seems to remarkably contain sufficient topological and/or morphological changes in order to drastically change the corresponding effective conductivities. Just for comparison, consider again the structures $\uuS^\mathrm{f}(1)$ and $\uuS^\mathrm{f}(2)$ illustrated in \autoref{fig_flipping} with corresponding effective conductivities for 32x resolution
\begin{equation}
	\bar\uuK^{\mathrm{f,32x}}(1)
	= 
	\begin{pmatrix}
	4.2363 & 0 \\
	0 & 1.8564
	\end{pmatrix}
	\ , \quad
	\bar\uuK^{\mathrm{f,32x}}(2)
	= 
	\begin{pmatrix}
	2.8439 & 0 \\
	0 & 2.0439
	\end{pmatrix}
	\ .
\end{equation}
The percolating structure $\uuS^\mathrm{f}(1)$ yields for 32x resolution a massive conductivity in $x_1$ direction and a corresponding relative deviation of 40.39 \% from $\bar\uuK^\mathrm{32x}(1)$. In comparison, the effective conductivity of $\uuS^\mathrm{f}(2)$ is only 3.32\% away from $\bar\uuK^\mathrm{32x}(1)$, i.e., much closer than $\bar\uuK^\mathrm{32x}(2)$. Hereby, one should note again that $\uuS(1)$, $\uuS^\mathrm{f}(1)$ and $\uuS^\mathrm{f}(2)$ have different 2PC.

\begin{figure}[H]
\centering
\includegraphics[scale=1.2]{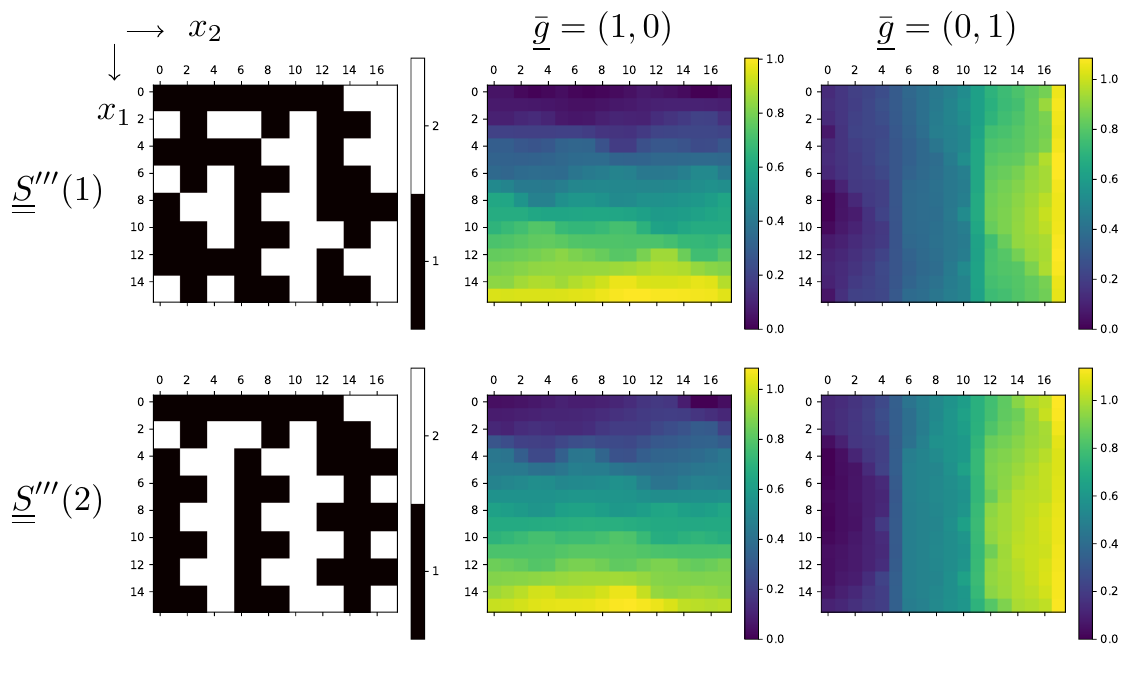}
\\
(a)
\\[2ex]
\includegraphics[scale=1.2]{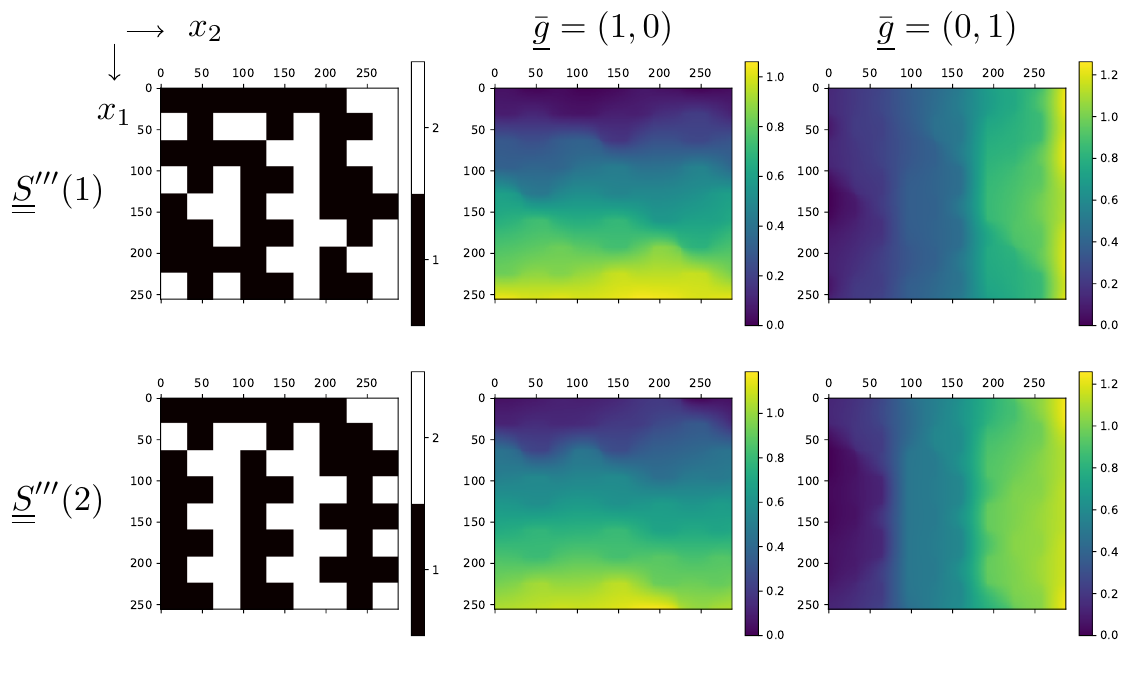}
\\
(b)
\caption{Structures $\uuS'''(1)$ and $\uuS'''(2)$ displayed in \autoref{fig_2d_example}: (a) with 2x resolution (left plots) and corresponding solution fields $u(\ux)$ for $\bar\ug = (1,0)$ (middle plots) and $\bar\ug = (0,1)$ right plots; (b) with 32x resolution (left plots) and corresponding solution fields $u(\ux)$ for $\bar\ug = (1,0)$ (middle plots) and $\bar\ug = (0,1)$ right plots}
\label{fig_h2_2}
\end{figure}

Of course, the magnitude of the deviation in the effective properties depend on the structure and material behavior taken into account. Consider as a second example the structures $\uuS'''(1) \sim \uuS'''(2)$ of \autoref{fig_2d_example}. These structures are generated from the previous $\uuS(1) \sim \uuS(2)$ by a phase-extension, kernel application and phase coalescence. The volume fraction of phase 1 changes to $v_1 = 41/72 \approx 57 \%$. For $k_2 < k_1$ and $v_1 > 1/2$ the generated structures $\uuS'''(1) \sim \uuS'''(2)$ can be interpreted as weakened structures of a material with conductivity $k_1$ for phase 1. Analogous increase in the resolution to 2x and 32x yields the results depicted in \autoref{fig_h2_2} (a) and (b), respectively. 

\begin{figure}[H]
\centering
\includegraphics[width=0.45\textwidth]{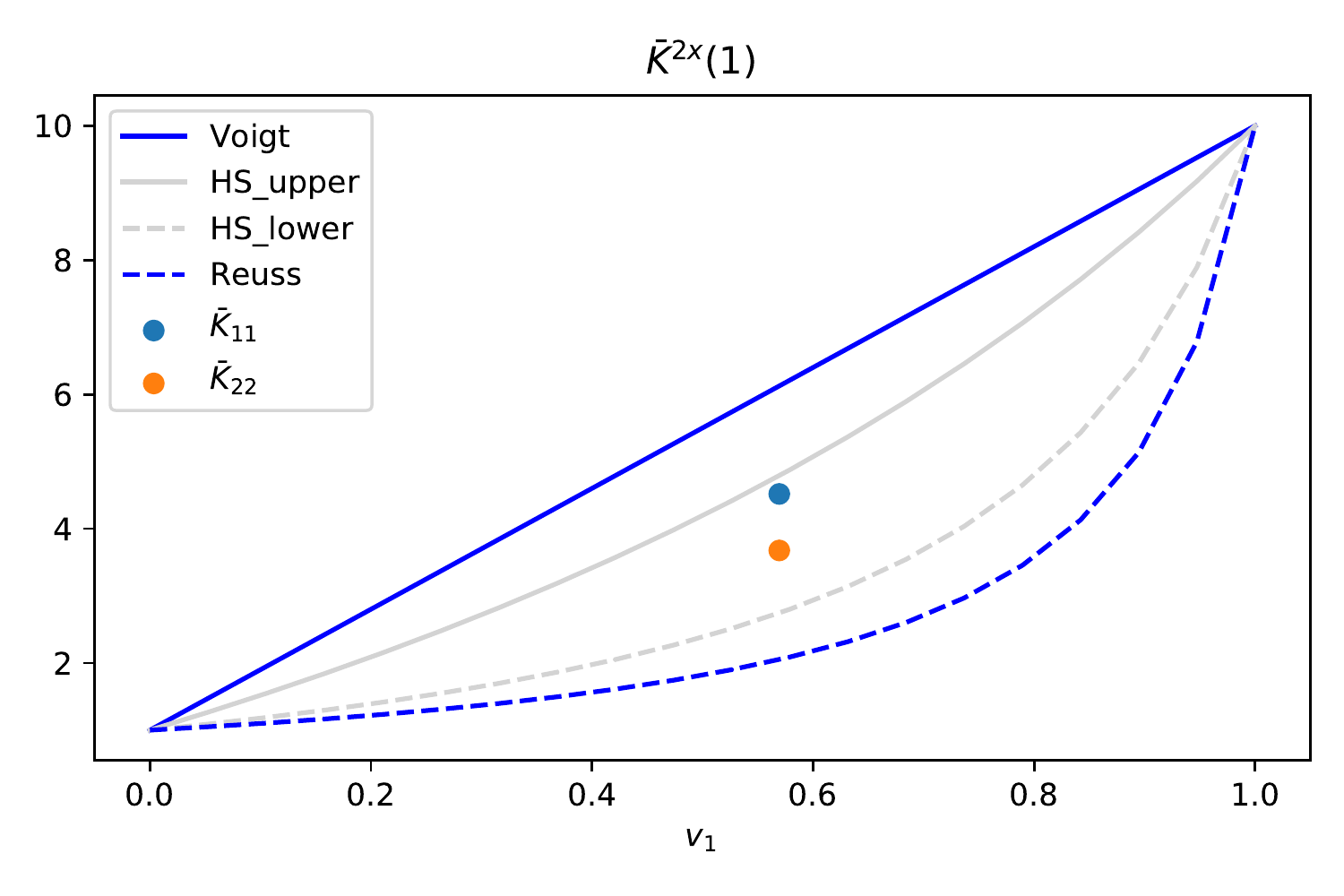}
\includegraphics[width=0.45\textwidth]{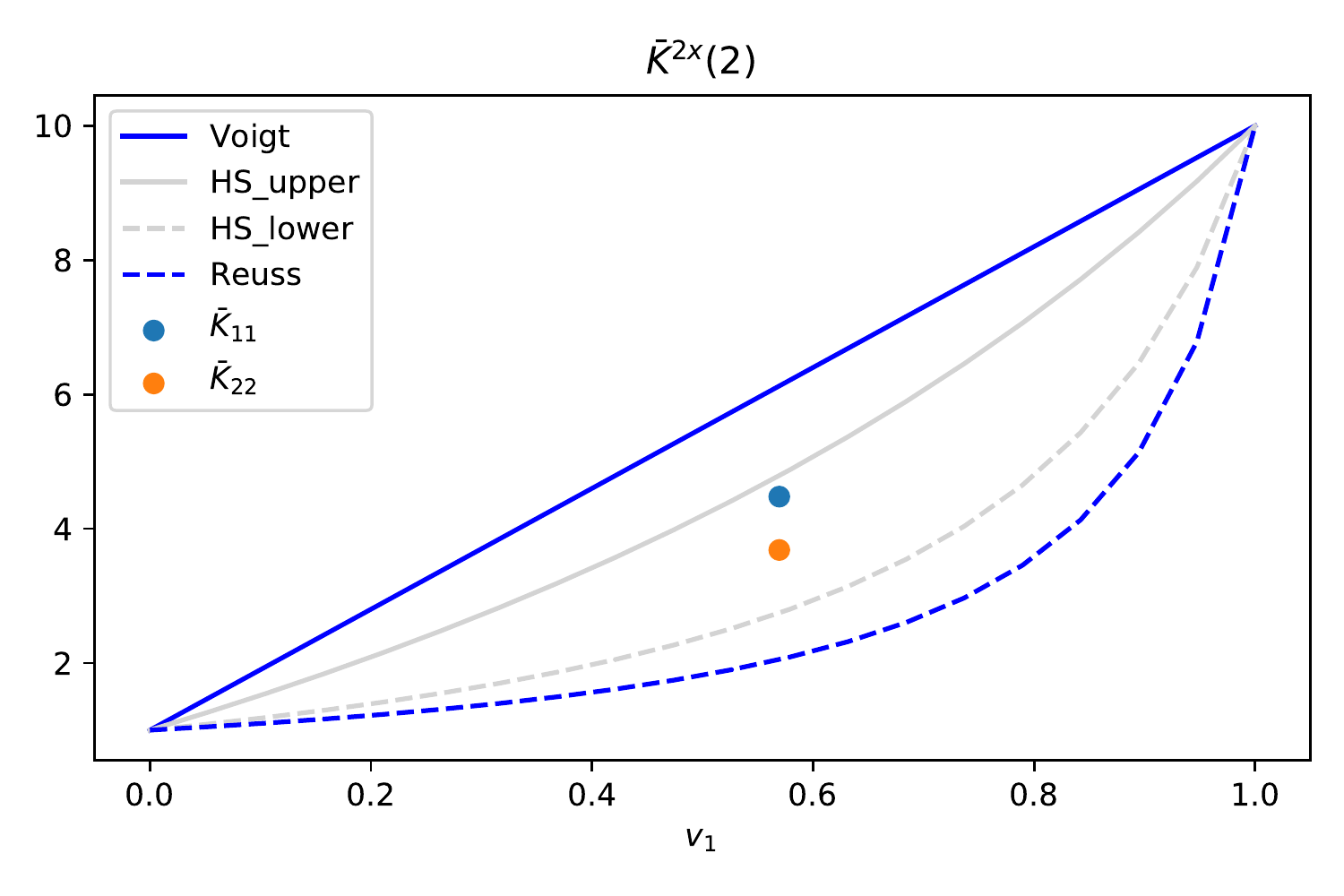} \\
\includegraphics[width=0.45\textwidth]{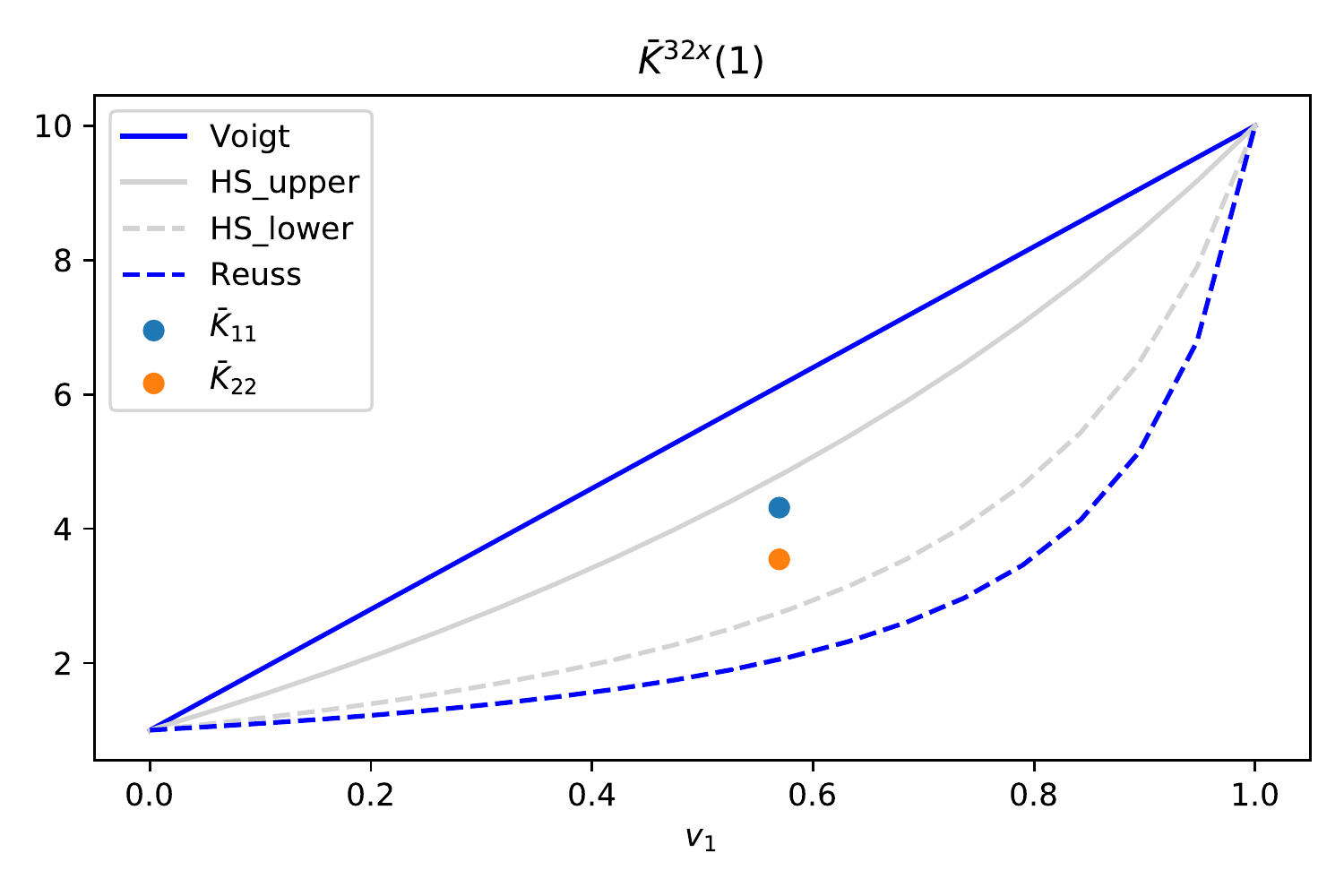}
\includegraphics[width=0.45\textwidth]{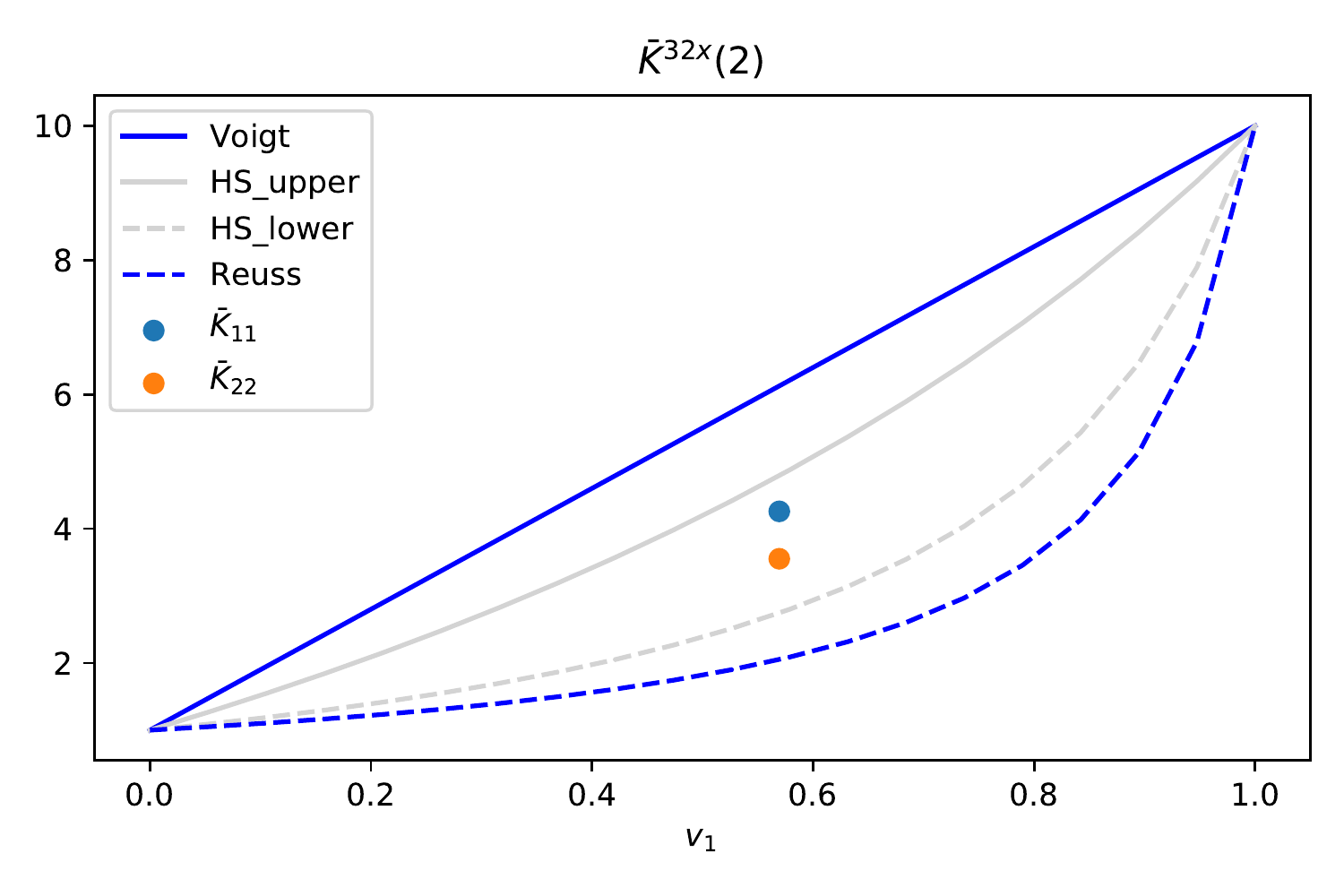}
\caption{Evaluation of Voigt, Reuss and Hashin-Shtrikman bounds for $v_1 \in [0,1]$ with $k_1 = 10$ and $k_2 = 1$; the effective conductivites of the structures $\uuS'''(1) \sim \uuS'''(2)$ for 2x and 32x resolution is depicted by the points at $v_1 \approx 0.57$: the $\bar{K}_{11}$ component of the corresponding effective conductivity is displayed by the corresponding blue point, while the $\bar{K}_{22}$ component is depicted by the corresponding orange point.}
\label{fig_bounds_2}
\end{figure}

The corresponding effective conductivities for $k_1 = 10$ and $k_2 = 1$ are
\begin{equation}
	\bar\uuK^{\text{2x}}(1)
	=
	\begin{pmatrix}
	4.5169 & 0.2396 \\
	0.2396 & 3.6756
	\end{pmatrix}
	\ , \quad
	\bar\uuK^{\text{2x}}(2)
	=
	\begin{pmatrix}
	4.4771 & 0.2304 \\
	0.2304 & 3.6825
	\end{pmatrix}
	\ .
\end{equation}
and
\begin{equation}
	\bar\uuK^{\text{32x}}(1)
	=
	\begin{pmatrix}
	4.3133 & 0.2890 \\
	0.2890 & 3.5431
	\end{pmatrix}
	\ , \quad
	\bar\uuK^{\text{32x}}(2)
	=
	\begin{pmatrix}
	4.2577 & 0.2859 \\
	0.2859 & 3.5518
	\end{pmatrix}
	\ .
\end{equation}
In \autoref{fig_bounds_2}, the Voigt, Reuss and Hashin-Shtrikman bounds are illustrated, together with the effective conductivities of the structures $\uuS'''(1)$ and $\uuS'''(2)$.

Evaluation of the analogous relative deviation yields
\begin{equation}
	\frac{\norm{\bar\uuK^{\text{2x}}(1) - \bar\uuK^{\text{2x}}(2)}}{\norm{\bar\uuK^{\text{2x}}(1)}}
	= 0.72 \%
	\ , \quad
	\frac{\norm{\bar\uuK^{\text{32x}}(1) - \bar\uuK^{\text{32x}}(2)}}{\norm{\bar\uuK^{\text{32x}}(1)}}
	= 1.00 \%
\end{equation}
such that this second example demonstrates that for some 2PC-equivalent structures the deviation in the effective conductivities may also be small. 

\section{Conclusions}
\label{sec_conclusions}

It has been shown for discrete periodic structures that when starting from given/root 2PC-equivalent structures, new 2PC-equivalent structures can be constructed systematically based on the basic operations
\begin{itemize}
\item phase extension through trivial embedding,
\item kernel-based extension, and
\item phase coalescence.
\end{itemize}
These results allow for the immediate generation of of an infinite number of 2PC-equivalent structures of arbitrary dimension, size, aspect ratio and number of phases through composition of the basic operations. Open-source software is provided for the generation of root 2PC-structures and construction of 2PC-equivalent child structures. With the provided software interested readers can generate their own root structures and derive 2PC-equivalent child structures for the application of interest as, e.g., for fiber- or particle-reinforced materials and polycrystals. Examples of 2PC-equivalent structures have been demonstrated and analyzed in the context of homogenization theory of periodic media for the two-dimensional conductivity problem. It has been shown that 2PC-equivalent structures exist, which may show significant or almost negligible deviations in the effective material behavior. The magnitude of the deviation depends not only on the structures, but also on the considered material behavior. The presented structures may serve as benchmark problems for homogenization schemes using the 2PC and possibly further microstructure information in order to test their accuracy or uncertainty with respect to the predicted effective properties. 


\section*{Acknowledgements}

This work is funded by the German Research Foundation (DFG) – project number 257987586, grant FR2702/6  within the Emmy-Noether program and project number 406068690, grant FR2702/8 within the Heisenberg program. Further, the project received funding by the Deutsche Forschungsgemeinschaft (DFG, German Research Foundation) under Germany‘s Excellence Strategy -- EXC-2075 -- project number 390740016.


\bibliography{library}


\begin{appendix}
\section{Auxiliary proofs based on DFT}
\label{app_proofs}

The property \eqref{dft_prop_zp} $\cF(\utA^{\{\uz\}}) = (\cF(\utA))^{[\uz]}$ connected to the case-sensitive trivial embedding \eqref{eq_zp_def1} - \eqref{eq_zp_def2}, and repetition operation \eqref{eq_rep_def1} - \eqref{eq_rep_def2} can be shown as follows. Consider $\uz \in \bbN^N$ and $\utA \in \bbC^{\uP}$ with $\uP \in \bbN^D$. For the case $N \leq D$ \eqref{dft_prop_zp} is show by
\begin{eqnarray}
	(\cF(\utA^{\{\uz\}}))_{p_1 \dots p_N p_{N+1} \dots p_D}
	&=& 
	\sum_{q_1 = 0}^{z_1P_1 - 1}
	\dots 
	\sum_{q_N = 0}^{z_NP_N - 1}
	\sum_{q_{N+1} = 0}^{P_{N+1} - 1}
	\dots
	\sum_{q_D = 0}^{P_D - 1}
	A^{\{\uz\}}_{q_1 \dots q_N q_{N+1} \dots q_D}
	\nonumber
	\\
	&&
	\exp\left(
	-i2\pi
	\left(
	\frac{p_1 q_1}{z_1 P_1}
	+ \dots
	+ \frac{p_N q_N}{z_N P_N}
	+ \frac{p_{N+1} q_{N+1}}{P_{N+1}}
	+ \dots
	+ \frac{p_D q_D}{P_D}
	\right)
	\right)
	\nonumber
	\\
	&=&
	\sum_{q_1 = 0}^{P_1 - 1}
	\dots 
	\sum_{q_N = 0}^{P_N - 1}
	\sum_{q_{N+1} = 0}^{P_{N+1} - 1}
	\dots
	\sum_{q_D = 0}^{P_D - 1}
	A_{q_1 \dots q_N q_{N+1} \dots q_D}
	\nonumber
	\\
	&&
	\exp\left(
	-i2\pi
	\left(
	\frac{p_1 q_1}{P_1}
	+ \dots
	+ \frac{p_N q_N}{P_N}
	+ \frac{p_{N+1} q_{N+1}}{P_{N+1}}
	+ \dots
	+ \frac{p_D q_D}{P_D}
	\right)
	\right)
	\nonumber
	\\
	&=&
	(\cF(\utA))_{p_1 \dots p_N p_{N+1} \dots p_D}
	\ .
\end{eqnarray}
For the case $N > D$
\begin{eqnarray}
	(\cF(\utA^{\{\uz\}}))_{p_1 \dots p_D p_{D+1} \dots p_N}
	&=& 
	\sum_{q_1 = 0}^{z_1P_1 - 1}
	\dots 
	\sum_{q_D = 0}^{z_DP_D - 1}
	\sum_{q_{D+1} = 0}^{z_{D+1} - 1}
	\dots
	\sum_{q_N = 0}^{z_N - 1}
	A^{\{\uz\}}_{q_1 \dots q_D q_{D+1} \dots q_N}
	\nonumber
	\\
	&&
	\exp\left(
	-i2\pi
	\left(
	\frac{p_1 q_1}{z_1 P_1}
	+ \dots
	+ \frac{p_D q_D}{z_D P_D}
	+ \frac{p_{D+1} q_{D+1}}{z_{D+1}}
	+ \dots
	+ \frac{p_N q_N}{z_N}
	\right)
	\right)
	\nonumber
	\\
	&=&
	\sum_{q_1 = 0}^{z_1P_1 - 1}
	\dots 
	\sum_{q_D = 0}^{z_DP_D - 1}
	\exp\left(
	-i2\pi
	\left(
	\frac{p_1 q_1}{z_1P_1}
	+ \dots
	+ \frac{p_D q_D}{z_DP_D}
	\right)
	\right)
	A^{\{\uz\}}_{q_1 \dots q_D 0 \dots 0}
	\nonumber
	\\
	&=&
	\nonumber 
	\sum_{q_1 = 0}^{P_1 - 1}
	\dots 
	\sum_{q_D = 0}^{P_D - 1}
	\exp\left(
	-i2\pi
	\left(
	\frac{p_1 q_1}{P_1}
	+ \dots
	+ \frac{p_D q_D}{P_D}
	\right)
	\right)
	A_{q_1 \dots q_D}
	\\
	&=&
	(\cF(\utA))_{p_1 \dots p_D}
\end{eqnarray}
holds, such that \eqref{dft_prop_zp} is then true for all cases. The property \eqref{dft_prop_zp2} $\cF(\utA^{\{\uz\}} \circledast \utB^{\{\uz\}}) = (\cF(\utA \circledast \utB))^{[\uz]} $ is derived based on \eqref{dft_prop_corr} and \eqref{dft_prop_zp} as follows
\begin{eqnarray}
	\cF(\utA^{\{\uz\}} \circledast \utB^{\{\uz\}})
	&=& 
	\overline{\cF\left(\overline{\utA}^{\{\uz\}}\right)}
	\odot
	\cF(\utB^{\{\uz\}})
	\nonumber
	\\
	&=&
	\overline{\cF(\overline{\utA})}^{[\uz]}
	\odot
	\cF(\utB)^{[\uz]}
	\nonumber
	\\
	&=&
	\left(
	\overline{\cF(\overline{\utA})}
	\odot
	\cF(\utB)
	\right)^{[\uz]}
	\nonumber
	\\
	&=&
	(\cF(\utA \circledast \utB))^{[\uz]}
	\ .
\end{eqnarray}

\section{Issues in \cite{Niezgoda2008}}
\label{app_niezgoda_counterexample}

For a compact notation in this appendix, we address the components of the 2PCs as $C_{\alpha\beta,\up} = C_{\alpha\beta,p_1 \dots p_D} \in \bbZ$ and the corresponding complex components of the DFT of the 2PC as $\hat{C}_{\alpha \beta,\up} \in \bbC$. Further, we define the number of events $I_{\alpha,\up} = 1$ in phase $\alpha$ as $\#_\alpha$
\begin{equation}
	\#_\alpha
	=
	\sum_{p_1 = 0}^{P_1 - 1}
	\dots
	\sum_{p_D = 0}^{P_D - 1}	
	I_{\alpha, \up}
	\ , \quad
	\#_\alpha \leq P_1 \dots P_D
	\ .
	\label{eq_app_ht}
\end{equation}
In \cite{Niezgoda2008} the following properties are presented:
\begin{align}
	\hat{C}_{\alpha \beta, \up}
	&= \overline{\hat{C}}_{\alpha \beta, (\uP - \up)}
	\ , \label{eq_app_niezgoda_1_dft_sym} \\
	\hat{C}_{\alpha \beta, \up}
	& = \overline{\hat{C}}_{\beta \alpha,\up}
	\ , \label{eq_app_niezgoda_2_2pc_def} \\
	\hat{C}_{\alpha \gamma,\up} \hat{C}_{\gamma \beta,\up}
	&= \hat{C}_{\gamma \gamma,\up} \hat{C}_{\alpha \beta, \up}
	\ , \label{eq_app_niezgoda_3_key} \\
	\sum_{\beta=1}^n \hat{C}_{\alpha \beta, \up}
	&= 
	\begin{cases}	
	(P_1 \dots P_D) \sqrt{\hat{C}_{\alpha \alpha, \underline{0}}} & \up = \underline{0} \ , \\
	0 & \text{else} \ ,
	\end{cases}
	\label{eq_app_niezgoda_4_2pc_last} \\
	\sum_{p_1 = 0}^{P_1 - 1}
	\dots
	\sum_{p_D = 0}^{P_D - 1}	
	\hat{C}_{\alpha \beta, \up}
	&= 
	\begin{cases}
	(P_1 \dots P_D) \sqrt{\hat{C}_{\alpha \alpha, \underline{0}}} & \alpha = \beta \ , \\
	0 & \text{else} \ ,
	\end{cases}
	\label{eq_app_niezgoda_5_finv_0}
	\\
	0 &\leq \hat{C}_{\alpha \beta, \underline{0}} \leq (P_1 \dots P_D)^2
	\label{eq_app_niezgoda_6_bound_0}
	\\
	0 &\leq |\hat{C}_{\alpha \beta, \underline{p}}| \leq (P_1 \dots P_D)^2
	\label{eq_app_niezgoda_7_bound_p}
	\ .
\end{align}
These properties are provided by the equations N.(15), N.(8), N.(9), N.(13), N.(14) and N.(11) in \cite{Niezgoda2008}. It is remarked that $\hat\uC_{\alpha \beta,\up}$ is related to $S \times {}^{np}F_\mathbf{k}$ in \cite{Niezgoda2008}, but with a different convention of the DFT, and, probably, the relation $|{}^{np}F_\mathbf{k}| \leq S^2$ in equation N.(11) is a typo and $|{}^{np}F_\mathbf{k}| \leq S$ was meant. The properties \eqref{eq_app_niezgoda_1_dft_sym} - \eqref{eq_app_niezgoda_7_bound_p} can easily be derived as follows. Property \eqref{eq_app_niezgoda_1_dft_sym} is the standard DFT symmetry of real signals. Property \eqref{eq_app_niezgoda_2_2pc_def} is obtained directly from the 2PC definition or based on \eqref{eq_2pc_gen_rel}. Property \eqref{eq_app_niezgoda_3_key} is considered a key result in \cite{Niezgoda2008} and can be obtained through simple rearrangement of the term $\bar{\hat{\utI}}_\alpha \odot \utilde{\hat{I}}_\gamma \odot \bar{\hat{\utI}}_\gamma \odot \utilde{\hat{I}}_\beta$. Property \eqref{eq_app_niezgoda_4_2pc_last} is exactly the determination of the last 2PC based on the $(n-1)$ counterparts, see \eqref{eq_2pc_last}, and is derived by taking the DFT of \eqref{eq_2pc_last} and exploiting $\hat{\utC}_{\alpha \alpha} = \bar{\hat{\utI}}_\alpha \odot \utilde{\hat{I}}_\alpha$ and $\hat{C}_{\alpha \alpha,\underline{0}} = \#_\alpha^2$. The property \eqref{eq_app_niezgoda_5_finv_0} is a reformulation of $C_{\alpha\beta, \underline{0}}$ expressed through the inverse DFT. Finally, the bounds \eqref{eq_app_niezgoda_6_bound_0} and \eqref{eq_app_niezgoda_7_bound_p} can be derived based on $|\hat{I}_{\alpha,\up}| \leq \#_\alpha$, \eqref{eq_app_ht} and the even tighter relation
\begin{equation}
	|\hat{C}_{\alpha \beta, \up}|
	\leq
	\#_\alpha \#_\beta
	= \hat{C}_{\alpha \beta, \underline{0}}
	\leq 
	(P_1 \dots P_D)^2
	\ .
	\label{eq_app_mybounds}
\end{equation}

In \cite{Niezgoda2008}, property \eqref{eq_app_niezgoda_3_key} is reformulated as
\begin{equation}
	\hat{C}_{\alpha \beta,\up}
	= \frac{\hat{C}_{\alpha \gamma,\up} \hat{C}_{\gamma \beta,\up}}{\hat{C}_{\gamma \gamma,\up}}
	\label{eq_app_niezgoda_wrong}
\end{equation}
without any remark that this is potentially only valid, if and only if $\hat{C}_{\gamma \gamma, \up} \neq 0$ for any $\up$. Still, \eqref{eq_app_niezgoda_wrong} is attractive since for given $\gamma$ and given 2PCs $\{\utC_{\gamma 1},\dots,\utC_{\gamma (n-1)}\}$, $\utC_{\gamma n}$ can be determined by \eqref{eq_2pc_last} (or equivalently \eqref{eq_app_niezgoda_4_2pc_last}) and \eqref{eq_app_niezgoda_wrong} could be used in combination with \eqref{eq_2pc_gen_rel} (or equivalently \eqref{eq_app_niezgoda_2_2pc_def}) in order to determine all remaining 2PCs - without any usage of \eqref{eq_app_niezgoda_1_dft_sym} or \eqref{eq_app_niezgoda_5_finv_0}. In \cite{Niezgoda2008} \eqref{eq_app_niezgoda_wrong} is considered as the cornerstone which implies the statement that for an $n$-phase structure if for any choice of $\gamma$ all corresponding 2PCs $\{\utC_{\gamma 1},\dots,\utC_{\gamma (n-1)}\}$ (or their DFTs) are known, then all 2PCs can be computed, i.e., $(n-1)$ 2PCs fully determine all $n^2$ 2PCs. This is stated without proof that the system of equations implicitly described through \eqref{eq_app_niezgoda_1_dft_sym} - \eqref{eq_app_niezgoda_7_bound_p} is uniquely solvable for given $\{\utC_{\gamma 1},\dots,\utC_{\gamma (n-1)}\}$ for any $\gamma$ or consideration of probably vanishing $\hat{C}_{\gamma \gamma, \up}$. Since no proper algebraic proof is provided in \cite{Niezgoda2008}, specially for cases with vanishing $\hat{C}_{\gamma \gamma,\up}$, the statement of \cite{Niezgoda2008} is considered here as a guess for special structures.

In order to explicitly show that cases with vanishing $\hat{C}_{\gamma \gamma,\up}$ exist, consider the following structure
\begin{equation}
	\uS = (1,1,1,2,2,3) \in \bbZ^6
	\label{eq_app_S}
\end{equation}
with $n=3$ phases and with indicators
\begin{align}
	\uI_1 &= (1,1,1,0,0,0) \ , \\
	\uI_2 &= (0,0,0,1,1,0) \ , \\
	\uI_3 &= (0,0,0,0,0,1) \ .
\end{align}
The 2PC for the current structure evaluate to 
\begin{align}
	\uC_{11} &= (3, 2, 1, 0, 1, 2) \ , &
	\uC_{12} &= (0, 1, 2, 2, 1, 0) \ , &
	\uC_{13} &= (0, 0, 0, 1, 1, 1) \ , \\
	\uC_{21} &= (0, 0, 1, 2, 2, 1) \ , &
	\uC_{22} &= (2, 1, 0, 0, 0, 1) \ , &
	\uC_{23} &= (0, 1, 1, 0, 0, 0) \ ,\\
	\uC_{31} &= (0, 1, 1, 1, 0, 0) \ , &
	\uC_{32} &= (0, 0, 0, 0, 1, 1) \ , &
	\uC_{33} &= (1, 0, 0, 0, 0, 0) \ .
\end{align}
A set of $(n-1)n/2 = 3$ independent 2PC for the current example is given by $\{\uC_{11},\uC_{12},\uC_{22}\}$. The corresponding DFT's of the 2PC $\uC_{\alpha \alpha}$ for $\alpha \in \{1,2,3\}$ are computed as
\begin{align}
	\hat\uC_{11} &= (9, 4, 0, 1, 0, 4) \ , \label{eq_app_dft_c11}\\
	\hat\uC_{22} &= (4, 3, 1, 0, 1, 3) \ , \label{eq_app_dft_c22}\\
	\hat\uC_{33} &= (1, 1, 1, 1, 1, 1) \ . \label{eq_app_dft_c33}
\end{align}
For the example structure \eqref{eq_app_S} with $n=3$ of this appendix, the DFTs of the 2PCs evaluate as given in \eqref{eq_app_dft_c11}, \eqref{eq_app_dft_c22} and \eqref{eq_app_dft_c33}, such that it becomes evident that some components of the respective DFTs can vanish and \eqref{eq_app_niezgoda_wrong} can \emph{not} be evaluated for all $p$. For example, if $\gamma=1$ is chosen and $\{\hat\uC_{11},\hat\uC_{12}\}$ are considered as given, then $\hat\uC_{13}$ can be determined through \eqref{eq_app_niezgoda_4_2pc_last}. Then, using \eqref{eq_app_niezgoda_2_2pc_def}, $\hat\uC_{21}$ and $\hat\uC_{31}$ can be computed. Subsequently, based on \eqref{eq_app_niezgoda_wrong} most components of $\hat\uC_{22}$ and $\hat\uC_{23}$ can be determined. Only $\hat{C}_{\alpha \beta,2}$ and $\hat{C}_{\alpha \beta,4}$ can \emph{not} be determined with \eqref{eq_app_niezgoda_wrong} since $\hat{C}_{11,2} = \hat{C}_{11,4} = 0$ hold. Only the additional consideration of \eqref{eq_app_niezgoda_1_dft_sym} and \eqref{eq_app_niezgoda_5_finv_0} allows the determination of all complex entries of all 2PCs for the present example. Changing to $\gamma=2$ does not entirely help since $\hat{C}_{22,3} = 0$ holds, only for the choice of $\gamma=3$ one can compute all 2PCs without \eqref{eq_app_niezgoda_1_dft_sym} or \eqref{eq_app_niezgoda_5_finv_0}. But this would have to be known \emph{a priori}. 

Naturally, higher-dimensional examples also exist, as the two-dimensional structure
\begin{equation}
	\uuS 
	=
\left(
\begin{array}{ccc}
 2 & 3 & 1 \\
 2 & 2 & 1 \\
 2 & 3 & 3 \\
 2 & 2 & 3 \\
\end{array}
\right)
	\in \bbN^{4 \times 3}
	\label{eq_app_S_2d}
\end{equation}
with $n=3$ and where $\{\uuC_{11},\uuC_{12},\uuC_{22}\}$ constitute a set of $(n-1)n/2 = 3$ independent 2PC, with DFTs 
\begin{equation}
	\hat\uuC_{11}
	=
	\left(
\begin{array}{ccc}
 4 & 4 & 4 \\
 2 & 2 & 2 \\
 0 & 0 & 0 \\
 2 & 2 & 2 \\
\end{array}
\right)
	\ , \quad
	\hat\uuC_{12}
	=
	\left(
\begin{array}{ccc}
 12 & -6-2 i \sqrt{3} & -6+2 i \sqrt{3} \\
 0 & 0 & 0 \\
 0 & 0 & 0 \\
 0 & 0 & 0 \\
\end{array}
\right)
	\ , \quad
	\hat\uuC_{22}
	=
	\left(
\begin{array}{ccc}
 36 & 12 & 12 \\
 0 & 0 & 0 \\
 4 & 4 & 4 \\
 0 & 0 & 0 \\
\end{array}
\right)
	\ .
\end{equation}
For $\hat{\uuC}_{11}$ and $\hat{\uuC}_{22}$, respectively, 3 and 6 entries vanish. This second example provides a counterexample for the claim of \cite{Niezgoda2008}. For the choice of $\gamma=1$, i.e., assuming given $\{\hat\uuC_{11},\hat\uuC_{12}\}$, one may attempt to solve for all $\hat\uuC_{\alpha \beta}$ but the system \eqref{eq_app_niezgoda_1_dft_sym} - \eqref{eq_app_niezgoda_7_bound_p} can not be solved \emph{uniquely}. The authors of the present work have been able to generate the following analytic deviation in DFT space 
\begin{equation}
	\uuDelta_{\alpha \beta}
	=
	\begin{cases}
\left(
\begin{array}{ccc}
 0 & 0 & 0 \\
 0 & 0 & 0 \\
 -2 & 1 & 1 \\
 0 & 0 & 0 \\
\end{array}
\right)
	& (\alpha,\beta) \in \{(2,2),(3,3)\}
		\\
\left(
\begin{array}{ccc}
 0 & 0 & 0 \\
 0 & 0 & 0 \\
 2 & -1 & -1 \\
 0 & 0 & 0 \\
\end{array}
\right)
	& (\alpha,\beta) \in \{(2,3),(3,2)\}
		\\
\left(
\begin{array}{ccc}
 0 & 0 & 0 \\
 0 & 0 & 0 \\
 0 & 0 & 0 \\
 0 & 0 & 0 \\
\end{array}
\right)
	& \text{else}
	\end{cases}
\end{equation}
such that the analytic $\hat\uuC_{\alpha \beta}$ from the original structure $\uuS$ given in \eqref{eq_app_S_2d} and the arrays $\hat\uuC'_{\alpha \beta} = \hat\uuC_{\alpha \beta} + \uuDelta_{\alpha \beta}$ with the choice of $\gamma=1$ fulfill $\hat\uuC_{\gamma \delta} = \hat\uuC'_{\gamma \delta} \ \forall \delta$, all equations in \eqref{eq_app_niezgoda_1_dft_sym} - \eqref{eq_app_niezgoda_5_finv_0} and the bounds \eqref{eq_app_mybounds}, such that the bounds \eqref{eq_app_niezgoda_6_bound_0} and \eqref{eq_app_niezgoda_7_bound_p} are automatically fulfilled. This finally implies that the claim of \cite{Niezgoda2008} that $(n-1)$ 2PCs uniquely determine all $n^2$ 2PCs based on \eqref{eq_app_niezgoda_1_dft_sym} - \eqref{eq_app_niezgoda_7_bound_p} is not true, in general. Whether $\hat\uuC'_{\alpha \beta}$ corresponds to a real structure or not, is debatable and is left open for future investigation, but the ambiguity of the system \eqref{eq_app_niezgoda_1_dft_sym} - \eqref{eq_app_niezgoda_7_bound_p} remains true for cases where $\hat{C}_{\gamma \gamma, \up} = 0$ holds. Based on the provided Python 3 software, the interested reader can check this analytic example, see \autoref{app_software}. It should be remarked that examples with vanishing DFT components in \emph{all} phases exist. For example, for the structure 
\begin{equation}
	\uuS = 
\left(
\begin{array}{ccc}
 1 & 1 & 3 \\
 1 & 2 & 3 \\
 1 & 1 & 3 \\
 3 & 1 & 2 \\
\end{array}
\right)
\end{equation}
with $n=3$, in $\hat{\uuC}_{11}$, $\hat{\uuC}_{22}$ and $\hat{\uuC}_{33}$, respectively, 2, 2 and 3 entries vanish. A final example for $n=4$ is given by
\begin{equation}
	\uuS =
	\left(
	\begin{array}{ccccc}
	 1 & 2 & 3 & 4 & 2 \\
	 1 & 4 & 4 & 4 & 4 \\
	 2 & 4 & 3 & 1 & 4 \\
	 4 & 4 & 1 & 4 & 4 \\
	\end{array}
	\right)
	\in \bbN^{4 \times 5}
\end{equation}
where for $\hat{\uuC}_{11}$, $\hat{\uuC}_{33}$ and $\hat{\uuC}_{44}$, respectively, 3, 10 and 4 entries vanish, and for which the authors have also not been able to solve uniquely for the choice $\gamma=1$.

The determination of the minimal number of independent 2PCs requires, from the perspective of the authors of the present work, further investigation. A proper algebraic proof based on constraints probably extending \eqref{eq_app_niezgoda_1_dft_sym} - \eqref{eq_app_niezgoda_5_finv_0} and \eqref{eq_app_mybounds} is still needed, especially in view of unknown number and positions of vanishing DFT components $\hat{C}_{\gamma \gamma,\up}$.

\section{Open-source Python 3 software}
\label{app_software}

Scientific work is expected to be transparent, reproducible and of practical use. This motivates the authors to offer a Python 3 implementation of the results of the present work through the Github repository \cite{Fernandez2020eq2pc_github}
\begin{center}
	\url{https://github.com/DataAnalyticsEngineering/EQ2PC}
\end{center}
The authors hope that the offered repository and the provided open-source software are of use for the interested reader. The open-source software (BSD 3-Clause License) contains a package with the main module
\begin{itemize}
\item \texttt{eq2pc.py} .
\end{itemize}
This module contains the main routines for the generation of root 2PC-equivalent structures and generation of derived ones based on the operations of this work (phase extension through trivial embedding, kernel application and phase concatenation). The auxiliary modules
\begin{itemize}
\item \texttt{tensor.py}
\item \texttt{database.py}
\item \texttt{conductivity.py}
\item \texttt{elasticity.py}
\end{itemize}
contain routines from tensor algebra (\texttt{tensor.py}), routines for the generation of an extendable \texttt{hdf5} database of structures with branch-based evaluation of derived structures (\texttt{database.py}), routines for the evaluation of two- and three-dimensional bounds of the effective conductivity of structures (\texttt{conductivity.py}), and routines for the evaluation of three-dimensional bounds of the effective elastic stiffness of structures (\texttt{elasticity.py}). 
The demonstration files 
\begin{itemize}
\item \texttt{demo1\_eq2pc.py}
\item \texttt{demo2\_bounds.py}
\item \texttt{demo3\_database.py}
\end{itemize}
are offered to show the respective functionalities. Further, the script
\begin{itemize}
\item \texttt{demo4\_niezgoda2008.py}
\end{itemize}
shows the analytic example discussed for \eqref{eq_app_S_2d} based on symbolic routines.
\end{appendix}

\end{document}